\documentclass[twocolumn,tighten]{aastex631}

\usepackage{makecell}  

\newcommand{\msun}{M_\odot}
\usepackage{amsmath}
\usepackage{enumitem}

\newcommand{\knnAcc}{96.0}
\newcommand{\knnBalAcc}{91.7}

\newcommand{\svcAcc}{98.3}
\newcommand{\svcBalAcc}{97.7}

\newcommand{\NNAcc}{98.5 \ensuremath{\pm} 0.4}
\newcommand{\NNBalAcc}{97.9 \ensuremath{\pm} 0.2}

\newcommand{\NNAccBest}{98.6}
\newcommand{\NNBalAccBest}{98.4}

\newcommand{\knnAbsMOne}{0.021}
\newcommand{\knnAbsMTwo}{$1 \times 10^{-5}$}
\newcommand{\knnRelMOne}{0.0037}
\newcommand{\knnRelMTwo}{0.011}

\newcommand{\svrAbsMOne}{0.028}
\newcommand{\svrAbsMTwo}{0.021}
\newcommand{\svrRelMOne}{0.0045}
\newcommand{\svrRelMTwo}{0.015}

\newcommand{\NNAbsMOne}{4.67\ensuremath{\pm} 0.23} 
\newcommand{\NNAbsMTwo}{0.00047\ensuremath{\pm} 0.00038}
\newcommand{\NNRelMOne}{1.09 \ensuremath{\pm} 0.05}
\newcommand{\NNRelMTwo}{1.76 \ensuremath{\pm} 0.15}

\newcommand{\NNAbsMOneBest}{4.41}
\newcommand{\NNAbsMTwoBest}{0.00037}
\newcommand{\NNRelMOneBest}{1.05}
\newcommand{\NNRelMTwoBest}{1.51}
\newcommand{\NNRelMOneBestPer}{0.11}%
\newcommand{\NNRelMTwoBestPer}{0.15}%

\newcommand{\moeAcc}{98.5\ensuremath{\pm} 0.1}
\newcommand{\moeBalAcc}{98.2\ensuremath{\pm} 0.2}
\newcommand{\moeAccBest}{98.6}
\newcommand{\moeBalAccBest}{98.5}

\newcommand{\moeAbsMOne}{12.5\ensuremath{\pm} 1.5} 
\newcommand{\moeAbsMTwo}{0.01\ensuremath{\pm} 0.01} 
\newcommand{\moeRelMOne}{3.1\ensuremath{\pm} 0.3} 
\newcommand{\moeRelMTwo}{8.4\ensuremath{\pm} 1.7} 

\newcommand{\moeAbsMOneBest}{10.4}
\newcommand{\moeAbsMTwoBest}{0.004}
\newcommand{\moeRelMOneBest}{2.7}
\newcommand{\moeRelMTwoBest}{7.5}
\newcommand{\collAIderAccTest}{98.2}
\newcommand{\collAIderAccInt}{100}
\newcommand{\collAIderAccExt}{90.6}
\newcommand{\collAIderAccTAMS}{94.9}

\newcommand{\collAIderBalAccTest}{98.2}
\newcommand{\collAIderBalAccInt}{100}
\newcommand{\collAIderBalAccExt}{83.8}
\newcommand{\collAIderBalAccTAMS}{89.4}

\newcommand{\collAIderAbsMOneTest}{0.0043}
\newcommand{\collAIderAbsMOneInt}{0.091}
\newcommand{\collAIderAbsMOneExt}{1.25}
\newcommand{\collAIderAbsMOneTAMS}{0.0086}

\newcommand{\collAIderAbsMTwoTest}{0.0}
\newcommand{\collAIderAbsMTwoInt}{0.0}
\newcommand{\collAIderAbsMTwoExt}{0.0}
\newcommand{\collAIderAbsMTwoTAMS}{0.0}

\newcommand{\collAIderRelMOneTest}{0.0011}
\newcommand{\collAIderRelMOneInt}{0.0022}
\newcommand{\collAIderRelMOneExt}{0.011}
\newcommand{\collAIderRelMOneTAMS}{0.0041}

\newcommand{\collAIderRelMOneExtPer}{1.1}%

\newcommand{\collAIderRelMTwoTest}{0.0014}
\newcommand{\collAIderRelMTwoInt}{0.0036}
\newcommand{\collAIderRelMTwoExt}{0.012}
\newcommand{\collAIderRelMTwoTAMS}{0.0044}

\newcommand{\collAIderRelMTwoExtPer}{1.2}%

\definecolor{darkgreen}{rgb}{0,0.6,0}

\shorttitle{}

\begin{document}

\title{Machine Learning Methods for Stellar Collisions. I. Predicting Outcomes of SPH Simulations}

\author[0000-0002-0933-6438]{Elena Gonz\'{a}lez Prieto}
\affil{Department of Physics \& Astronomy, Northwestern University, Evanston, IL 60208, USA}
\affil{Center for Interdisciplinary Exploration \& Research in Astrophysics (CIERA), Northwestern University, Evanston, IL 60201, USA}
\affil{NSF-Simons AI Institute for the Sky (SkAI), 172 E. Chestnut St., Chicago, IL 60611, USA}

\author[0000-0002-7444-7599]{James C.\ Lombardi, Jr.}
\affiliation{Department of Physics, Allegheny College, Meadville, Pennsylvania 16335, USA}

\author[0000-0003-0984-4456]{Sanaea C.\ Rose}
\affil{Center for Interdisciplinary Exploration \& Research in Astrophysics (CIERA), Northwestern University, Evanston, IL 60201, USA}
\affil{NSF-Simons AI Institute for the Sky (SkAI), 172 E. Chestnut St., Chicago, IL 60611, USA}

\author[0009-0003-8690-8297]{Charles F.A.\ Gibson}
\affil{Department of Physics \& Astronomy, Northwestern University, Evanston, IL 60208, USA}
\affil{Center for Interdisciplinary Exploration \& Research in Astrophysics (CIERA), Northwestern University, Evanston, IL 60201, USA}
\affil{NSF-Simons AI Institute for the Sky (SkAI), 172 E. Chestnut St., Chicago, IL 60611, USA}

\author[0000-0003-3987-3776]{Christopher E.\ O'Connor}
\affil{Center for Interdisciplinary Exploration \& Research in Astrophysics (CIERA), Northwestern University, Evanston, IL 60201, USA}
\affil{NSF-Simons AI Institute for the Sky (SkAI), 172 E. Chestnut St., Chicago, IL 60611, USA}

\author[0000-0003-2539-8206]{Tjitske Starkenburg}
\affil{Department of Physics \& Astronomy, Northwestern University, Evanston, IL 60208, USA}
\affil{Center for Interdisciplinary Exploration \& Research in Astrophysics (CIERA), Northwestern University, Evanston, IL 60201, USA}
\affil{NSF-Simons AI Institute for the Sky (SkAI), 172 E. Chestnut St., Chicago, IL 60611, USA}

\author[0000-0003-4412-2176]{Fulya K{\i}ro\u{g}lu}
\affil{Center for Interdisciplinary Exploration \& Research in Astrophysics (CIERA), Northwestern University, Evanston, IL 60201, USA}
\affil{NSF-Simons AI Institute for the Sky (SkAI), 172 E. Chestnut St., Chicago, IL 60611, USA}

\author[0000-0002-4086-3180]{Kyle Kremer}
\affil{Department of Astronomy \& Astrophysics, University of California, San Diego; La Jolla, CA 92093, USA}

\author[0009-0004-5934-9650]{Tristan C.\ Sand}
\affiliation{Department of Physics, Loyola University Chicago, Chicago, IL 60660, USA}

\author[0000-0002-7132-418X
]{Frederic A.\ Rasio}
\affil{Department of Physics \& Astronomy, Northwestern University, Evanston, IL 60208, USA}
\affil{Center for Interdisciplinary Exploration \& Research in Astrophysics (CIERA), Northwestern University, Evanston, IL 60201, USA}
\affil{NSF-Simons AI Institute for the Sky (SkAI), 172 E. Chestnut St., Chicago, IL 60611, USA}

\begin{abstract}
Stellar collisions can occur frequently in dense cluster environments and play a crucial role in producing exotic phenomena, from blue stragglers in globular clusters to high-energy transients in galactic nuclei. Successive collisions and mergers of massive stars could also lead to the formation of massive black holes, serving as seeds for supermassive black holes in the early Universe. While analytic fitting formulae exist for predicting collision outcomes, they do not generalize across different energy scales or stellar evolutionary phases. Smoothed particle hydrodynamics (SPH) simulations are often used to compute the outcomes of stellar collisions, but even at low resolution, their computational cost makes running on-the-fly calculations during an $N$-body simulation quite challenging. Here, we present a new grid of $27,720$ SPH calculations of main-sequence star collisions, spanning a wide range of masses, ages, relative velocities, and impact parameters. Using this grid, we train machine learning models to predict both collision outcomes (merger versus disruption or flyby) and final remnant masses. We compare the performance of 
nearest neighbors, support vector machines, and neural networks, achieving classification balanced accuracy of \NNBalAccBest$\%$, and regression relative errors as low as \NNRelMOneBestPer$\%$ and \NNRelMTwoBestPer$\%$ for the final stars $1$ and $2$, respectively. We make our trained models publicly available as part of the package \texttt{collAIder}, enabling rapid predictions of stellar collision outcomes in $N$-body models of dense star cluster dynamics.

\end{abstract}

\section{Introduction}

Stellar collisions and mergers are important across a wide range of dense stellar systems, from galactic nuclei to open clusters \citep{Spitzer1966,Sanders1970, Lombardi+96, Hurley2001, Shara2002, Hurley2005, Glebbeek2008, Dale2009, Balberg2023, Gonzalez_Prieto2024}. The properties and outcomes of these collisions vary drastically depending on the environment. In galactic nuclei, stars collide at relative velocities of hundreds to thousands of kilometers per second  \citep[e.g.,][]{Rauch1999,Genzel+10, Rose+23}, while in globular clusters, typical velocity dispersions are only tens of kilometers per second \citep[e.g.,][]{Pryor1993, Harris2010, Baumgardt2018}. Whether a collision results in a merger, mass stripping, or complete disruption depends on this relative velocity, as well as the pericenter distance and the mass, radius, and evolutionary stage of each star \citep[e.g.,][]{Freitag2005}.

Understanding how these dynamical parameters shape collision outcomes has proven crucial for explaining exotic stellar objects and phenomena. For example, blue stragglers, stars that appear on the main sequence (MS) beyond the turnoff, could form through collisions between lower-mass MS stars \citep[e.g.,][]{Leonard1989,Lombardi+96, Sills2009}. Stellar collisions have also been studied for their possible role as gravitational-wave progenitors \citep[e.g.,][]{Zwart2004, Gurkan2006, Kremer2020, Seoane2023, Gonzalez_Prieto2024, kiroglu2025}. Transient events are also expected, such as tidal disruption events \citep[TDEs; e.g.,][]{Perets2016, Kremer2019b, Lopez2019, Wang2021, Kremer2022, Ryu2022, Kiroglu2023, Xin2024, Ryu2024, Rose2025}, energetic explosions from collisions between hypervelocity stars in galactic nuclei \citep{Balberg2013, Brutman2024, Hu2024}, and collision-driven events \citep{vanderMerwe2024, vanderMerwe2025, Peng2025}. Furthermore, high-speed collisions between intermediate-mass stars can produce stripped stellar remnants \citep[e.g.,][]{Lai1993, Rauch1999, Freitag2007, Rose+23} with abundance ratios similar to those observed in the TDE  ASASSN-14ko \citep[e.g.,][]{Payne2021,Gibson+24}. Lastly, mergers can also produce exotic signatures, such as gamma-ray production from a black hole merger with a helium star \citep[e.g.,][]{Connor2025, Neights2026} or luminous red novae \citep[e.g.,][]{Ivanova2013, MacLeod2017, Metzger2017, Kirilov2025}.

Currently, most $N$-body models of dense stellar systems have simplified treatments for stellar collisions, such as the ``sticky sphere'' approximation, in which no mass is lost during the collision. While this assumption is adequate for nearly parabolic collisions in globular clusters, hyperbolic collisions in nuclear star clusters can produce significant shock-driven mass loss. Accurately predicting the detailed outcomes of stellar collisions, such as remnant masses and structures, therefore requires three-dimensional hydrodynamics simulations that can resolve relevant thermodynamical and hydrodynamic processes, using methods such as smooth particle hydrodynamics \citep[SPH;][]{Rasio1991, Monaghan1992, Gaburov2010} 
or grid-based and moving-mesh methods \citep{Arepo}. However, these detailed simulations are computationally expensive, sometimes requiring days of wall-clock time per collision, making on-the-fly calculations in $N$-body cluster simulations infeasible. As a result, a practical approach is to precompute large grids of SPH results and develop fitting formulae \citep[e.g.,][]{Lai1993,Rauch1999, Rose+25}. 

Machine learning (ML) offers a unique opportunity to overcome these computational limitations. Using traditional ML algorithms, a model can be trained on a large grid of simulations to learn the mapping between inputs and outputs. This approach has been proven effective in the context of planetary collisions. For example, \cite{Cambioni2019} compared classification and regression performance across multiple algorithms. This work was extended to predict core mass fractions \citep{Cambioni2021} and orbital parameters \citep{Emsenhuber2020}. Recently, \cite{Seoane2025} and \cite{Rose+25} applied similar techniques in the context of stellar collisions, with the former using an older grid of stellar collision SPH simulations presented in \cite{Freitag2005} and the latter using a new grid of equal-mass but higher-resolution collision calculations. 

For this study, we compute a comprehensive new grid of SPH results for stellar collisions, spanning wide ranges of ages, stellar masses, pericenter distances, and relative velocities. We improve upon the SPH simulations presented in \cite{Freitag2005} by using accurate \texttt{MESA} stellar profiles and exploring the effects of stellar age on collision outcomes. Following the methodology of \cite{Cambioni2019}, we compare the performance of multiple ML algorithms on both classification and regression tasks. The classification task predicts the number of surviving stars, while the regression task estimates their final masses.  Furthermore, we compare different ML methods, including nearest neighbors \citep{Cover1967}, support vector machines \citep{SVM}, and neural networks \citep[e.g.,][]{McCulloch, Hornik, LeCun2015}. 


In Section~\ref{sec:methods}, we describe our SPH simulations, including the \texttt{MESA} stellar models used as initial conditions. In Sections~\ref{sec:classification} and~\ref{sec:regression}, we present our ML framework for classifying stellar collision outcomes and predicting the properties of final remnants. In Section~\ref{sec:MoE} we compare these models against a Mixture of Experts architecture. We then introduce \texttt{collAIder} in Section~\ref{sec:collAIder}, a new software that leverages the trained ML models to make rapid predictions for stellar outcomes. Finally, in Section~\ref{sec:stresstests} we evaluate model performance on interpolated and extrapolated data. We summarize our results and outline future applications in Section~\ref{sec:discussion}.

\section{Methods}
\label{sec:methods}

\subsection{Simulation Grid}
\label{sec:grid}
Our training set of SPH results is built by sampling across five parameters: the age of the system ($t$), the primary mass ($M_1$), the secondary mass ($M_2$), the velocity at infinity ($v_\infty$), and the distance of closest approach (pericenter distance, $r_{\rm p}$). Consistent with assumptions made in many star cluster modeling codes, we assume that all stars are coeval. However, we account for the mass-dependent timescale for stars to contract onto the MS, and we collide only MS stars. For example, at an age of 1~Myr, we only perform collisions involving the most massive stars in the grid, since stars below $4\msun$ are still in the pre-MS stage. Furthermore, some of the lower-mass stars will not reach the zero-age main sequence (ZAMS) before the higher-mass stars evolve off the MS. As a result, we do not compute collisions between the highest- and lowest-mass stars in the sampled range. We restrict this initial grid to stars only on the MS because stellar structure varies greatly across the full age range of the sampled stellar masses. Accurately modeling the evolution of a wider range of collision outcomes across time would require a very finely sampled grid of pre-MS, MS, and giant star encounters, an undertaking which we reserve for future work.

For the base grid, ages are sampled logarithmically in gigayears at $(-3, -2.5, -2, -1.5, -1, -0.5, 0, 0.5, 1)$, together with an additional point at $13.7\,$Gyr  (Hubble time). Stellar masses are sampled at $10$ values from $0.2$ to $64 \msun$. The sampled velocities are (10, 100, 250, 500, 1000, 2000, 4000, 8000, 16000)~km/s, chosen to span speeds typical of stellar collisions from globular clusters to galactic nuclei. Pericenter distances are sampled based on fractional enclosed mass $f$. For each collision, we choose $f$ (uniformly in steps of 0.1 from $0$ to $0.9$, with additional samples at $0.95$, $0.97$, $0.99$, and $1$), compute the radii $r_1(f)$ and $r_2(f)$ enclosing this mass fraction in each parent star, and set the pericenter distance to $r_{\rm p} = r_1(f) + r_2(f)$. For example, $f=0$ corresponds to a head-on collision, while $f=1$ corresponds to a grazing encounter at the sum of the two stellar radii. 

\begin{figure*}
    \centering
    \includegraphics[width=\linewidth]{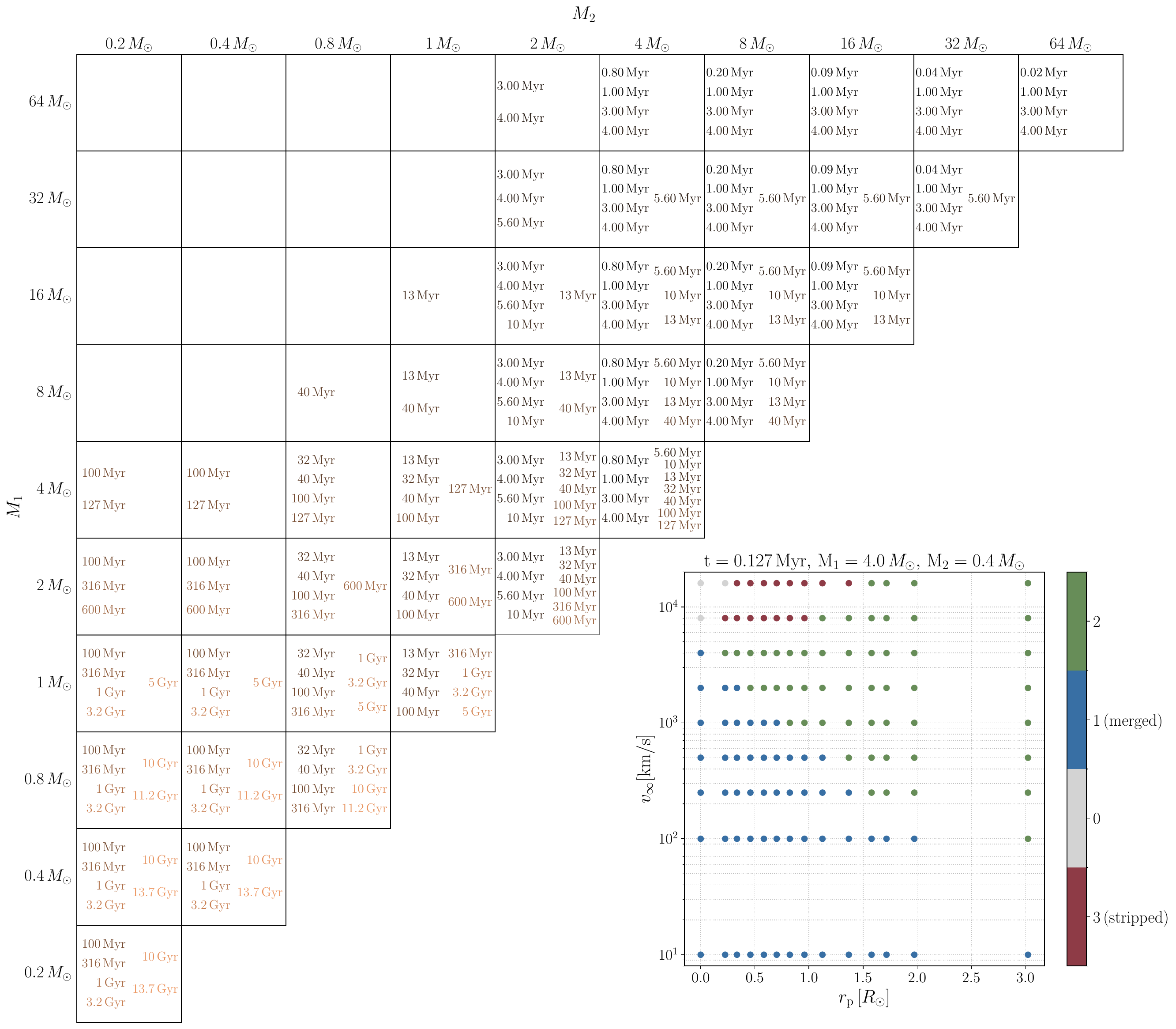}
    \caption{\label{fig:grid} Complete grid of SPH stellar collision simulations. The left and top axes show the primary and secondary masses involved in the collisions, respectively. Within each square, the corresponding collision times are shown, color-coded by age. In the lower-right corner, we show an example of how the pericenter distance ($r_p$) and velocity at infinity ($v_{\infty}$) are sampled for each combination of age, $M_1$, and $M_2$ dimensions. The colors indicate the number of stellar remnants, distinguishing mergers from cases that result in a stripped star.}
    
\end{figure*}

We complement the base grid by sampling the terminal-age main sequence (TAMS) for all stellar masses, defining the TAMS as the point at which the central hydrogen mass fraction drops below $10^{-5}$. We also compute a grid at the ZAMS age for the $64, 32, 16, 8,$ and $4 \msun$ stars, where ZAMS is defined as the local minimum in luminosity at which nuclear burning provides more than $50\%$ of the luminosity. For the ZAMS grid, each star on the ZAMS is collided with all other stars that are on the MS at that time, although we do not simulate collisions among the other MS stars themselves. This brings the total grid size to $27{,}720$ SPH simulations of stellar collisions. The complete grid is shown in Figure \ref{fig:grid}, where the lower-right panel illustrates how each mass and time slice is sampled in $r_p$ and $v_{\infty}$. 

\subsection{Stellar Profiles}
To generate the realistic stellar profiles needed for the stellar collision simulations, we use \texttt{MESA} \citep[v24.08.1;][]{Paxton2011, Paxton2013, Paxton2015, Paxton2018, Paxton2019, Jermyn2023}. We scale the initial chemical composition of our stellar models relative to the protosolar helium abundance, $Y_\odot = 0.2703$ and metallicity, $Z_\odot = 0.0142$ of \citet{Asplund2009}. We adopt a metallicity of $Z = 0.01~Z_\odot$ and calculate the corresponding helium abundance using Equation (2) in \cite{MIST} (assuming $Y_{\rm p} = 0.249$ from \cite{Planck2016}), yielding a value of $Y = 0.249$. We treat the microphysical processes relevant to MS stars following the \texttt{Posydon} model grids \citep{Fragos2023, Andrews2025}, unless otherwise specified. All \texttt{MESA} inlists used to generate our stellar models are publicly available.~\footnote{The data are available on Zenodo: 
\dataset[doi:10.5281/zenodo.19392209]{
https://doi.org/10.5281/zenodo.19392209}}

We do not include stellar winds or rotation since we consider only collisions involving MS stars. At this metallicity and evolutionary stage, mass loss is not expected to be very efficient. Future grids extending beyond stars on the MS will incorporate stellar rotation and metallicity-dependent wind prescriptions.

\subsection{Smoothed Particle Hydrodynamics}
The stellar collisions~\footnote{The data are available on Zenodo: 
\dataset[doi:10.5281/zenodo.19615605]{
https://doi.org/10.5281/zenodo.19615605}} are performed using \texttt{StarSmasher} \citep{Rasio1991, Gaburov2010}, a Lagrangian SPH code in which each fluid particle is assigned a mass, position, velocity, and specific internal energy. The particles are evolved using a variational SPH formulation, while gravitational forces and energies are computed through direct summation on NVIDIA GPUs \citep{Gaburov2010}.  We neglect radiative cooling and heating.
Each particle has an extended density profile described by a Wendland C4 kernel \citep{Wendland1995} with a compact support $2h_i$ (where $h_i$ is the smoothing length of the particle). Artificial viscosity is computed as in \cite{Hwang2015}.  

Because all parent stars are on the MS, we employ an analytic equation of state that includes both ideal gas and radiation pressure.  Radiation pressure can contribute appreciably in our high-mass models and may become more important in shock-heated regions during collisions.
For each SPH particle, given its density $\rho$, specific internal energy $u$, and mean molecular mass $\mu$ (with dimensions of mass), we obtain the temperature $T$ by solving $u=\tfrac{3}{2}\,kT/\mu + aT^4/\rho$, where $k$ is Boltzmann’s constant and $a$ is the radiation constant. The resulting quartic equation in $T$ is solved analytically following \citet{Lombardi2006}. We then compute the pressure from $P=\rho kT/\mu + \tfrac{1}{3}aT^4$.

Each star is composed of $10{,}000$ particles with $\sim50$ neighbors, where the one-dimensional \texttt{MESA} stellar profiles are converted into three-dimensional SPH models by placing particles in a stretchy hexagonal close-packed (HCP) lattice, as described in Appendix A of \cite{Gibson+24}. The value of \texttt{equalmass} is set to $0.5$ for most models, producing unequal-mass particles with a number density $n \propto \sqrt\rho$ that distributes more particles near the stellar center and better resolves the core. After constructing the stellar profile, the particles are relaxed into hydrostatic equilibrium as described in \cite{Lombardi2006} and \cite{Gaburov2010}.  

For later postprocessing, we store snapshots of the system at an output interval roughly equal to the dynamical timescale of the more massive star:
$\Delta t_{\rm out}\equiv t_{\rm dyn}=\sqrt{R_{\rm max}^3/(G M_{\rm max})}=\sqrt{M_{\rm max}}$ in code units
($G=M_\odot=R_\odot=1$), where for the purposes of assigning a $\Delta t_{\rm out}$ value, we adopt the approximate mass--radius relation
$R/R_\odot=(M/M_\odot)^{2/3}$.  Here, $R_{\rm max}$ and $M_{\rm max}$ are the radius $R$ and mass $M$, respectively, of the more massive of the two parent stars. We compute the bound mass of each star at every snapshot using an iterative energy-based procedure similar to that of \citet{Lombardi2006}. At a given snapshot, we first compute the center-of-mass (COM) and preliminary mass of each stellar component. We then evaluate the specific mechanical energy of each SPH particle with respect to each star’s COM. 

This mechanical energy includes only kinetic and gravitational potential terms, as recommended by \cite{Nandez2014}.
A particle is considered gravitationally bound to star $j$ if its mechanical energy relative to that star’s COM is negative. In particular, for particle $i$ relative to component $j$, we require \mbox{$v_{ij}^2/2 - G(M_j-m_i)/d_{ij} < 0$}, where $v_{ij}$ is the particle’s speed in the frame of star $j$’s COM, 
$d_{ij}$ is its distance from that center, $m_i$ is the mass of the particle, and $M_j$ is the current mass of star $j$. This approach is equivalent to using the Bernoulli equation at late times, since internal energy and enthalpy become negligible in the outflow after sufficient adiabatic expansion.
To avoid spuriously identifying poorly resolved clumps as stars, we classify a bound component as a stellar remnant only if it contains at least $20$ SPH particles. Otherwise, its particles are reassigned to the unbound mass.

We iterate this procedure to refine the bound masses self-consistently. Based on its assignment from the previous snapshot (or the known configuration at $t=0$), each particle is tentatively assigned to the component (star 1 or star 2) for which it has negative energy, or to the unbound mass otherwise. If a particle has a negative mechanical energy with respect to both stars, it is assigned to whichever star yields the most negative value. We then update the COM and mass of each star based on these assignments and recompute the particle energies, repeating the procedure until no particle changes affiliation (typically converging within a few iterations). Particles that do not have negative mechanical energy with respect to either star are assigned to the unbound component.

The stopping time ($t_f$) is dynamically determined by evaluating the state of the system once per output interval, $\Delta t_{\rm out}$. At each interval, {\tt StarSmasher} identifies the current bound components and computes their orbital elements. If a merger is detected (i.e., if the number of bound components decreases), or if a binary remains bound, we extend the run by resetting $t_f=\max(t_f,\,t+30\,\Delta t_{\rm out})$, where $t$ is the current time, ensuring at least $30$ additional outputs. Whenever $t_f$ is updated, the total internal energy is stored. If this quantity later changes by more than $1\%$, the run is again extended by $30\,\Delta t_{\rm out}$ to avoid stopping during ongoing thermal readjustment.  In cases where fewer than $100$ particles remain bound to a remnant, we enforce a minimum integration time by setting $t_f=\max(t_f,\,100\,\Delta t_{\rm out})$ to ensure that the remnant is long-lived. The simulation ends when the time reaches $t=t_f$.

\section{Classification of Collision Outcomes}\label{sec:classification}

The first quantity we aim to predict is the qualitative outcome of the collision, which is closely related to the number of stars remaining after the encounter. In terms of survivors, there are three possible outcomes: $0$, $1$, or $2$ remnants. At sufficiently large pericenter distances, collisions are grazing enough that both stars survive, though not completely intact. As the pericenter distance decreases and collisions become more head-on, the outcome depends increasingly on the relative velocity, mass ratio, and stellar structure of the colliding stars. For any two colliding stars at sufficiently low relative velocities, collisions generally result in mergers. However, at high relative velocities and small pericenter distances, collisions can become energetic enough to unbind material from both stars, leaving no surviving remnants. 

A special scenario arises in unequal-mass collisions where, at sufficiently high velocities and moderately off-axis trajectories, one star is destroyed while the other survives. In these cases, the surviving star is stripped of its envelope, and we therefore refer to such events as the ``stripping'' scenario. Although this outcome leaves the same number of surviving stars as a merger, we treat it as a separate class because the underlying hydrodynamic evolution and the properties of the final remnant are qualitatively different. 


To distinguish between merger and stripped-star outcomes, we monitor the relative orbit and proximity of the two stellar components throughout the collision. During postprocessing, a merger is identified when the two stellar components become gravitationally bound in an encounter sufficiently tight that the semimajor axis of their orbit falls below a critical threshold related to their tidal radii and sizes. Specifically, the code estimates each star’s tidal disruption radius as $r_{{\rm tidal},i}=R_i\left(M_j/M_i\right)^{1/3}$, where $j\neq i$ and with the effective radius of component $i$ estimated as $R_i=(I_i/M_i)^{1/2}$. Here, $I_i$ is the moment of inertia about the rotation axis, $M_i$ is the total bound mass of component $i$, and $M_j$ is the total bound mass of the companion component (i.e., $j=2$ if $i=1$, while $j=1$ if $i=2$). If the semimajor axis falls below min($r_{{\rm tidal},i}, R_j)$ for either component, we consider this a merger and immediately combine the bound particle sets of the two stars. If no merger occurs and only one star remains, we classify the surviving object as a stripped star.


We therefore define a four-class classification task with labels ${0, 1, 2, 3}$, where classes $0$, $1$, and $2$ indicate the number of surviving stars and class $3$ represents a stripped-star outcome rather than a merger. Before training a model on the simulation data, we transform the input parameters so that they all span comparable ranges, thus facilitating model training. We apply variable-specific transformations to each component of the five-dimensional input space to account for their different dynamics ranges: 
\begin{equation} \label{eq:input}
\bigl\{ \log(t + 0.001), \log(r_{\rm p} + 0.1), \log(v_{\infty} + 10), \ln(M_1), \ln(M_2)\bigr\}
\end{equation}

where $t$ is measured in units of gigayears, $r_{\rm p}$ in solar radii, $v_{\infty}$ in kilometers per second, and the stellar masses in solar masses. To avoid singularities near zero, we introduce offsets for $t$, $r_{\rm p}$, and $v_{\infty}$ with the same units and scales as the corresponding variables. 

In the context of stellar collisions, the pericenter distance is often normalized by the stellar radii to obtain a dimensionless parameter. However, this introduces a strong sensitivity to errors in the stellar radius estimates, which can be significant in some cases--- for example, for massive stars in the Single Star Evolution (\texttt{SSE}) code \citep[see Figure 8 in][]{Agrawal2020}. To avoid this issue, we train our ML model using pericenter distances expressed in units of solar radii. Consequently, the range of pericenter distances in the training set varies in scale depending on the stellar masses involved. 


In addition, all methods presented below standardize the data according to

\begin{equation} \label{eq:norm}
x'_i = \frac{x_i - \mu_i}{\sigma_i}.
\end{equation}
where $\mu_i$ and $\sigma_i$ are the mean and standard deviation of input parameter $i$, respectively. 

The dataset is divided into $70\%$ training, $15\%$ validation, and $15\%$ testing, corresponding to $19,404$, $4158$, and $4158$ samples, respectively. Using stratified sampling, we ensure that each split preserves the overall label proportions. The dataset exhibits a class imbalance for the stripping (label $3$) and mutually destructive (label $0$) cases. For instance, in the training set, these classes represent only $4.8\%$ and $8.1\%$ of the samples, respectively. We therefore prioritize models that optimize predictions across all classes, rather than optimizing performance on the overrepresented classes. Below, we present the three algorithms used for the classification task and compare their performance.

\vspace{0.5cm}
\subsection{k-Nearest Neighbors} \label{sec:class_$k$-NN}

$k$-nearest neighbors ($k$-NN) \citep{Cover1967} is a well-established ML algorithm that classifies data points based on the labels of their ``$k$'' closest neighbors. The number of neighbors considered, the distance metric, and the weighting scheme are all hyperparameters that can be tuned to optimize model performance. We perform a grid search over the number of neighbors (from $3$ to $25$), the distance metric (Euclidean or Manhattan), and the weighting scheme (uniform for all points or weighted by inverse distance). Performance is evaluated using balanced accuracy \footnote{Defined as $\frac{1}{k} \sum_{i=1}^{k} \frac{\mathrm{TP}_i}{\mathrm{TP}_i + \mathrm{FN}_i}$ where $k$ is the number of classes, TP is the true positives, and FN the false negatives.}, which averages the classification accuracy across classes and reduces sensitivity to class imbalance. The set of hyperparameters that yields the highest performance consists of five neighbors, Euclidean distance, and a weighting scheme in which points are weighted inversely proportional to their distance. The resulting model achieves a test balanced accuracy of \knnBalAcc \%. Both the accuracy and balanced accuracy are reported in Table~\ref{table:class_stats}, alongside the performance of the ML methods described below. 

\begin{deluxetable}{c | c | c }
\tabletypesize{\scriptsize}
\setlength{\tabcolsep}{0.7\tabcolsep}  
\centering
\tablecaption{\label{table:class_stats}Classification Performance}  
\tablehead{
    \colhead{Method} &
    \colhead{Accuracy } &
    \colhead{Balanced Accuracy} \\
    \colhead{} &
    \colhead{(\%)} &
    \colhead{(\%)}} 
\startdata
$k$-NN  & \knnAcc & \knnBalAcc \\ 
SVC     &  \svcAcc & \svcBalAcc \\  
NN      &  \makecell{\NNAcc \\ (\text{Best: } \NNAccBest)} 
        &  \makecell{\NNBalAcc \\ (\text{Best: } \NNBalAccBest)} 
\enddata
\tablecomments{Performance on the classification task across different methods. For the neural network (NN), we report the mean and standard deviation across $10$ runs with optimized hyperparameters and different random seeds. The best-performing run (shown in parentheses) is used in subsequent figures.}
\end{deluxetable}

\subsection{Support Vector Machines} \label{sec:class_SVM}

Support vector machines \citep[SVMs;][]{SVM} are supervised ML algorithms that determine decision boundaries by finding hyperplanes that best separate different classes in kernel space. Through kernel transformations, SVMs can classify nonlinear data, making them particularly effective for complex datasets and multiclass classification problems. 

As in the previous section, we perform a grid search to find the best-performing set of hyperparameters using the support vector classifier (SVC) available in \texttt{scikit-learn} \citep{scikit-learn}. The optimized parameters and explored values include ($1$) the kernel used to transform the input data into a higher-dimensional feature space (Gaussian radial basis function, \texttt{rbf}, or polynomial); ($2$) \texttt{C}, the regularization parameter controlling the complexity of the decision boundaries ($0.1, 1, 10, 100, 1000, 10000$); and ($3$) \texttt{gamma}, which controls how much influence a single data point has ($0.001, 0.01, 0.1, 1$). Within the polynomial kernel function, we additionally search over varying degrees ($2$ and $3$). We also apply class weights to account for the class imbalance in the dataset.

As expected, we find \texttt{rbf} to be the optimal kernel, as it offers greater flexibility in nonlinear decision boundaries. The best-performing hyperparameters are \texttt{C}=$10,000$ and \texttt{gamma}$ = 0.1$, yielding a test balanced accuracy of \svcBalAcc$\%$. Since the optimized value of \texttt{C} lies at the edge of the explored parameter range, we perform an additional localized search at higher values; however this yields an optimized value of \texttt{C}=$10,000$.

\subsection{Neural Network} \label{sec:class_NN}

Neural networks (NNs) learn mappings between inputs and outputs through a series of weighted transformations and nonlinear activation functions \citep[e.g.,][]{McCulloch, LeCun2015}. Our NN is a multilayer perceptron \citep[MLP;][]{Hornik} composed of three hidden layers with $512, 256,$ and $ 128$ neurons, respectively. Each hidden layer is followed by a rectified linear unit (ReLU) activation function \citep[e.g.,][]{fukushima1969visual,glorot2011deep,maas2013rectifier,Agarap2018} and layer normalization \citep{layer_norm}. 

All hyperparameters, including network architecture, batch size, optimizer, initial learning rate, and scheduler parameters, were optimized using a hyperparameter search with the Weights and Biases platform\footnote{https://wandb.ai/site/}. We performed a Bayesian optimization \citep{Snoek2012} sweep, which efficiently identifies hyperparameter combinations that improve model performance based on previous attempts. Table~\ref{table:class_NN_params} lists the ranges explored for each hyperparameter and their optimized values.

\begin{deluxetable}{c | c | c }
    \tabletypesize{\scriptsize}
    \setlength{\tabcolsep}{0.7\tabcolsep}  
    \centering
    \tablecaption{\label{table:class_NN_params} Classification Hyperparameters}  
    \tablehead{
    	\colhead{Parameter} &
    	\colhead{Optimized Value}  &
        \colhead{Search Range} } 
    \startdata
    $\rm Batch \, size$  &  $64 $ & $64 - 256 $ \\  
    $\rm Epochs$  &  $  500 $ & $500-2000 $ \\ 
    $\rm Optimizer$  &  $ \texttt{AdamW} $ & $\texttt{AdamW, sgd}$ \\ 
    $\rm Scheduler$  &  $ \texttt{CA-LR} $ & $\texttt{CA-LR, RLRP}$ \\  
    $\rm Learning \, rate$  &  $ 0.0002 $ & $10^{-5} - 0.1$ \\  
    $\rm \texttt{patience}$  &  $ \cdots $ & $10 - 60$ \\ 
    $\rm \texttt{factor}$  &  $ \cdots $ & $0.3 - 0.5$ \\  
    \enddata
    \tablecomments{Hyperparameter optimization search for the classification task. Acronyms: stochastic gradient descent (\texttt{sgd}), AdamW optimizer (\texttt{AdamW}), cosine annealing learning rate scheduler (\texttt{CA-LR}), and reduce learning rate on plateau scheduler (\texttt{RLRP}).}
\end{deluxetable}

The training and validation data are split into batches of size $64$. We use the AdamW \citep{Adamw} optimizer with a weighted cross-entropy loss function, where class weights account for the imbalanced representation of collision outcomes. The learning rate is dynamically adjusted using a \texttt{CosineAnnealingLR} scheduler, a feature available in the \texttt{PyTorch} library \citep{PyTorch}. The learning rate starts at $0.0002$ and decreases following a cosine function to a minimum value of $10^{-8}$. The model is trained for $500$ epochs, which we find sufficient for convergence. The \texttt{patience} and \texttt{factor} parameters belong to the \texttt{ReduceLROnPlateau} scheduler, where the learning rate decreases by a factor if a target metric (in our case the validation balanced accuracy) does not improve for a certain number of consecutive epochs (controlled by the \texttt{patience} parameter). No optimized values for these two hyperparameters are shown, since the best-performing model uses a \texttt{CosineAnnealingLR} scheduler. 

For each training run, we save the epoch with the highest validation balanced accuracy. Among all models in the hyperparameter optimization sweep, we select the one achieving the best validation balanced accuracy. We then asses its stability by performing an additional sweep across $10$ different random seeds. The mean and standard deviation of the test accuracy and balanced accuracy across these ten runs are reported in Table~\ref{table:class_stats}. We do not perform this stability test for the other methods, since $k$-NN is a nonparametric method and SVC finds a global solution, unlike neural networks where random weight initialization can lead to variability in model performance. The best-performing model from this sweep is used in all subsequent analyses and achieves a balanced accuracy of \NNBalAccBest$\%$.

\subsection{Algorithm Performance Comparison}

Table~\ref{table:class_stats} lists the test accuracies and balanced accuracies across the three methods. The SVC and NN achieve comparable performance, with both outperforming $k$-NN, which serves as our baseline. The strong performance of the SVC is expected, as it explicitly maximizes the margin to the decision boundary, which proves effective for our dataset.

\begin{figure*}
    \centering
    \includegraphics[width=\linewidth]{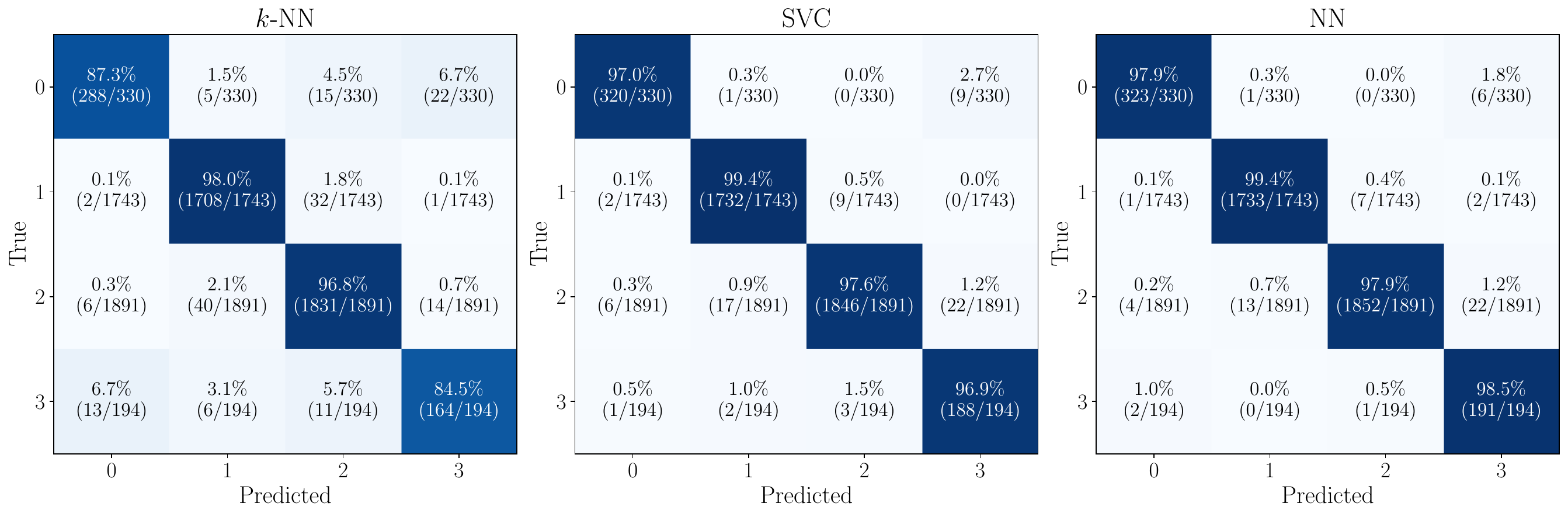}
    \caption{\label{fig:confusion_matrix} From left to right, confusion matrices for the test data using k-nearest neighbors algorithm ($k$-NN), support vector classifier (SVC), and a neural network (NN). The accuracy and total number of samples in each class are shown. }
    
\end{figure*}

\begin{figure*}
    \centering
    \includegraphics[width=\linewidth]{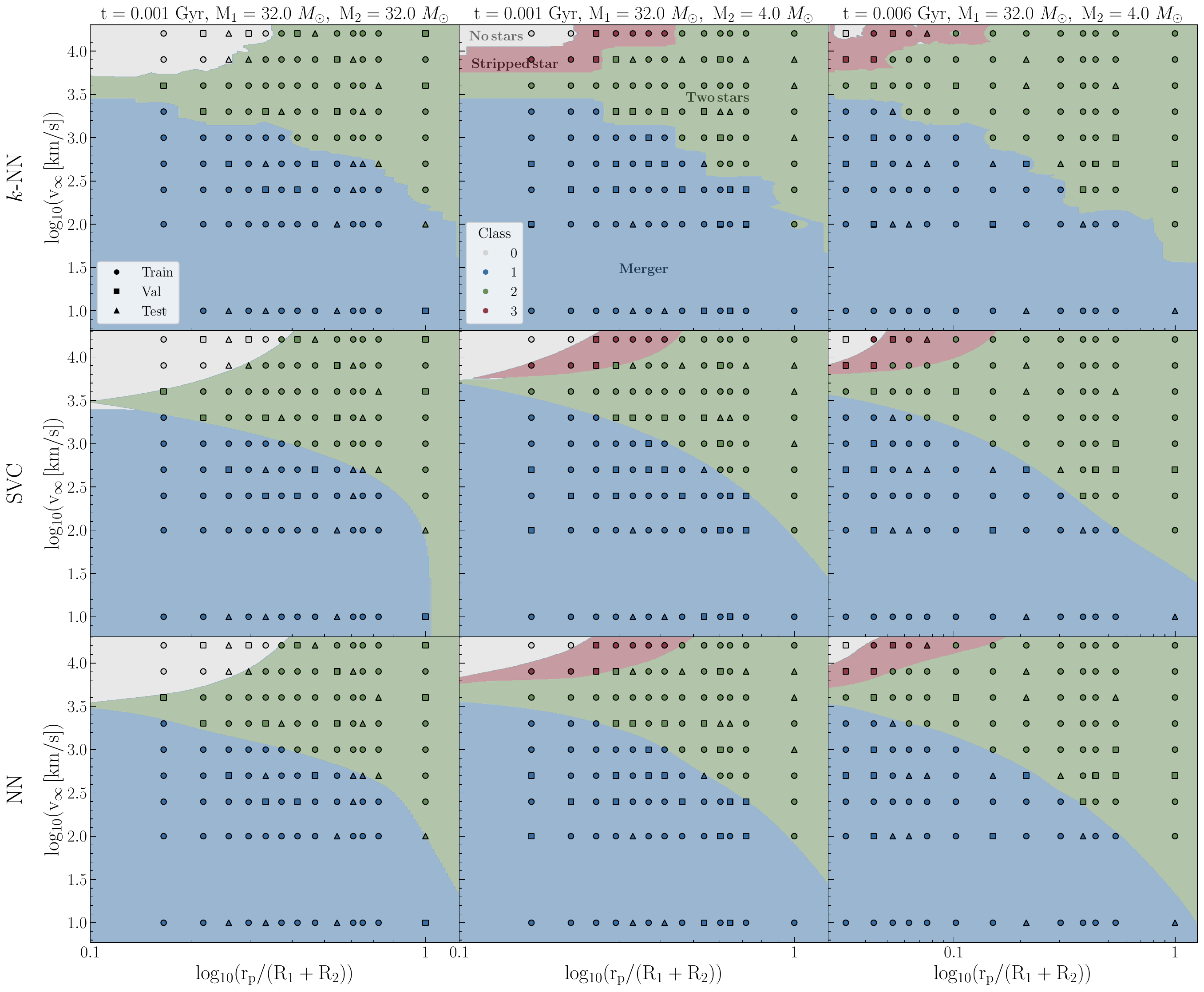}
    \caption{\label{fig:decision_boundaries} Two-dimensional examples of decision boundaries for the stellar collision outcome classification task as a function of pericenter distance ($r_p$) and speed at infinity ($v_\infty)$ for fixed stellar age and mass. Each color indicates a different class, where $0$ corresponds to mutual destruction, $1$ to mergers, $2$ to flybys, and $3$ to stripped stars. The different marker shapes indicate the training (circles), validation (squares), and test (triangles) datasets, and are colored according to the ground truth. Classification predictions are shown as background contours. Each column represents a different collision regime, where the first column shows an equal-mass collision, the middle panel a collision with an unequal mass ratio, and the third panel a system at the TAMS of the primary star. The x-axis is normalized to standardize the scales across columns.}
    
\end{figure*}

Figure~\ref{fig:confusion_matrix} shows the confusion matrices for each method, summarizing classification performance across all class labels. High values along the diagonal indicate optimal performance, corresponding to correct predictions for each class. Both the SVC and NN outperform $k$-NN across all classes. The $k$-NN and SVC methods perform best for classes with larger numbers of training samples and struggle with undersampled cases, particularly scenarios where no stars survive (class $0$) or where one stripped star remains after a destructive collision at high relative velocity (class $3$). For both methods, class $3$ exhibits the lowest accuracy. The reduced performance for class $3$ can be attributed to a combination of two factors: ($1$) its limited number of samples and ($2$) its strong dependence on stellar structure. In particular, the location and extent of this outcome region occurs vary significantly with evolutionary time and mass ratio. For example, as illustrated in Figure~\ref{fig:decision_boundaries}, the class $3$ region for a $32 \msun$ star at the TAMS stage (rightmost panel) occurs at lower pericenter distances than at earlier times (middle panel) and is absent in the equal-mass collision case (leftmost panel). 

The NN exhibits more balanced performance, partly due to the use of a weighted cross-entropy loss function that penalizes misclassifications in class $3$ more heavily. As a result, the NN outperforms the SVC on minority-class predictions. For both methods, the most common misclassifications occur when samples from label $2$ (a majority class) are classified as label $3$ (a minority class). This is also likely due to the chosen optimization strategy, which prioritizes minority-class performance at the expense of majority-class accuracy. 

To address class imbalance, we evaluate all models using balanced accuracy. In addition, the SVC and NN use class weighting, which penalizes misclassifications of minority classes more strongly during training. A similar weighting scheme is not available for $k$-NN, which likely contributes to its poor performance in minority classes as shown in Figure~\ref{fig:confusion_matrix}. Data augmentation methods such as random oversampling or SMOTE \citep{Chawla_2002} can also be used to address class imbalance. However, these methods must be applied with caution to high-dimensional datasets. In particular, synthetic data generated using SMOTE have been shown not to accurately represent the underlying distribution and therefore to decrease performance \citep[e.g.,][]{Blagus2013SMOTE}. Given that our class weighting and balanced accuracy scoring achieve strong performance and avoid potentially introducing incorrect synthetic samples, we choose not to include data augmentation in this work.

To visualize how these methods differ in their classification strategies, we show example decision boundaries in feature space in Figure~\ref{fig:decision_boundaries}. Each row showcases the predictions from the different ML algorithms, while each column represents different physical scenarios. 

The first column shows collision outcomes for an equal-mass interaction as a function of normalized pericenter distance and relative velocity at infinity. For visualization purposes, the pericenter distance has been normalized with respect to the sum of the stellar radii; however, the models were trained using the unnormalized pericenter values. Model predictions are shown as a shaded background, while the ground-truth labels are indicated by the colors of the data points. The middle column illustrates a collision between unequal-mass stars with a mass ratio of $\sim 0.13$. The rightmost column shows collision outcomes when the primary star, in this case a $32 \msun$ star, is on the TAMS. 

Because the pericenter distance is sampled across radii enclosing a fixed mass fraction, the pericenter distances at the TAMS are smaller than at earlier evolutionary stages. This reflects the structural evolution of the $32 \msun$ star, which develops a steeper density profile as it approaches the TAMS. Consequently, the third column highlights the sensitivity of the collision outcome to stellar structure, as the stripping class occurs at smaller relative pericenter distances compared to earlier evolutionary stages. 

Across each column, we show wo-dimensional projections of the decision boundaries produced by each algorithm. The $k$-NN method does not predict smooth decision boundaries, but rather has artificial features due to the intrinsic distances between the sampled neighbors. In contrast, the SVC produces smoother decision boundaries by finding hyperplanes that optimize the classification performance in kernel space. However, discontinuities appear at small pericenter distances, which are likely artifacts of projecting a higher-dimensional decision surface onto two dimensions and suggest that the SVC may produce overly complex decision surfaces that introduce nonphysical features. 

Despite the comparable test balanced accuracies for the SVC and NN, the smoothest and most physically consistent decision boundaries are achieved with the NN. This reflects the NN's ability to learn complex, nonlinear mappings between inputs and outputs, enabling it to capture dependencies across features and therefore produce more physically consistent classification predictions. In contrast, SVCs are optimized to find hyperplanes that separate classes using support vectors and therefore struggle in areas where data may be sparse or classes are imbalanced. 

As expected, most misclassifications occur near decision boundaries, where small changes in pericenter distance or velocity can yield different outcomes. The boundaries shift with stellar mass and age, highlighting the complexity of the classification task. Nevertheless, despite a high-dimensional dataset that spans a wide range in its feature space, both the SVC and NN achieve remarkable accuracies. In future work, we will incorporate active learning to guide the optimal expansion of the dataset and further improve model performance. Determining which input feature is most important is not straightforward, as the sensitivity of the collision outcome to a given input feature is not constant and depends on the values of all other features. For example, the age of the stars becomes more important in determining the collision outcome as the stars approach the TAMS, but has less influence than $r_\mathrm{p}$ and $v_{\infty}$ during MS evolution, as shown in Figure~\ref{fig:decision_boundaries}.

\section{Regression of Collision Outcome Properties} \label{sec:regression}

An additional set of key quantities to predict are the final properties of stellar collision remnants, in particular their final masses. This is motivated in part by the fact that most $N$-body codes---especially those modeling environments with low velocity dispersions---assume no mass loss during stellar mergers (the ``sticky sphere approximation''). However, this assumption can break down at small pericenter distances and higher velocities. 

To predict the final masses of the colliding stars (or the merger product, if the stars merge), we again compare three methods: $k$-NN, SVM, and NN. Rather than directly predicting the final masses, we predict three normalized mass fractions: 

\begin{equation} \label{eq:norb}
\biggl\{\frac{M_{1,\rm f}}{M_{\mathrm{tot},\rm i}}, \frac{M_{2,\rm f}}{M_{\mathrm{tot},\rm i}}, \frac{M_{\rm u, f}}{M_{\mathrm{tot},\rm i}}\biggr\}
\end{equation}
where $ M_{\rm 1,f}$ and $ M_{\rm 2,f}$ are the final masses of stars $1$ and $2$, respectively, $ M_{\rm tot,i}$ is the total initial mass of the system, and $ M_{\rm u,f}$ is the final unbound mass.

By construction, these three mass fractions always sum to unity due to mass conservation. Formatting the regression targets in this way encourages our models to predict physically consistent mass estimates. To reduce inhomogeneities in the dataset without altering the outcome of the collisions, we assign the final merger product mass to star $1$ in the case of a single remnant. The same input data are used as in Section~\ref{sec:classification}, and results are summarized in Table~\ref{table:reg_stats}. 

\subsection{k-Nearest Neighbors} \label{sec:reg_$k$-NN}

To perform regression using the $k$-NN method, we conduct the same search strategy as in Section~\ref{sec:class_$k$-NN} but evaluate models using the mean absolute error (MAE). Because the regression targets are mass fractions, the MAE is relative to the initial total mass of the system. To evaluate model performance on the test dataset, we multiply the predicted fractions by $ M_{\rm tot,i}$ and compute the absolute errors in the final masses, as well as relative errors for cases in which the star survives. The optimized $k$-NN model uses three neighbors, the Euclidean metric, and inverse-distance weighting. The resulting median absolute errors in $M_{\rm 1,f}$ and $M_{\rm 2,f}$ are \knnAbsMOne $\msun$ and \knnAbsMTwo $\msun$, respectively.

\subsection{Support Vector Regression} \label{sec:reg_SVR}

To perform regression with an SVM, we use epsilon-support vector regression \citep[SVR;][]{Vapni2000}. The algorithm learns a regression hyperplane, where $\epsilon$ defines the margin around the hyperplane, known as the $\epsilon$-insensitive tube. Training points within the margins do not contribute to the loss, while points outside the margins become support vectors that shape and optimize the hyperplane. Hyperparameter optimization is performed for each regression quantity separately, beginning with a coarse grid search and subsequently refining it. Full details of the hyperparameter search are outlined in Appendix~\ref{sec:svr_gridsearch}, including the choice of kernel, $\epsilon$, $\gamma$, and the regularization parameter. The optimal configuration for all regression targets uses an \texttt{rbf} kernel with $\texttt{C} = 1$, $\epsilon = 0.00001$, and $\gamma = 10$. The resulting best model achieves median absolute errors in $ M_{\rm 1,f}$ and $ M_{\rm 2,f}$ of \svrAbsMOne $\msun$ and \svrAbsMTwo $\msun$, respectively; additional performance metrics are listed in Table~\ref{table:reg_stats}. 

\begin{deluxetable}{lcccc}
\tabletypesize{\scriptsize}
\tablecaption{Regression Performance Metrics\label{table:reg_stats}}
\tablehead{
    \colhead{Method} &
    \multicolumn{2}{c}{Median Absolute Error ($\mathrm{M}_\odot$)} &
    \multicolumn{2}{c}{Median Relative Error}\\
    \colhead{} &
    \colhead{$M_{1,f}$} &
    \colhead{$M_{2,f}$} &
    \colhead{$M_{1,f}$} &
    \colhead{$M_{2,f}$}
}
\startdata
$k$-NN & \knnAbsMOne & \knnAbsMTwo & \knnRelMOne & \knnRelMTwo \\
SVR & \svrAbsMOne & \svrAbsMTwo & \svrRelMOne & \svrRelMTwo \\
NN ($\times10^{-3}$) & \makecell{\NNAbsMOne \\ (\text{Best:} \NNAbsMOneBest)} 
                           & \makecell{\NNAbsMTwo \\ (\text{Best:} \NNAbsMTwoBest)} 
                           & \makecell{\NNRelMOne  \\ (\text{Best:} \NNRelMOneBest)} 
                           & \makecell{\NNRelMTwo \\ (\text{Best:} \NNRelMTwoBest)} 
\enddata
\tablecomments{Regression performance metrics for k-nearest neighbors ($k$-NN), support vector regression (SVR), and neural networks (NN). We report the mean and standard deviation across $10$ runs with optimized hyperparameters and different random seeds. Fractional relative errors are computed only for cases in which the star survives. Note that performance metrics for the NN are scaled by a factor of $10^{-3}$.}
\end{deluxetable}

The median absolute errors for $ M_{\rm 1,f}$ are comparable between the $k$-NN and SVR models. However, this not the case for $ M_{\rm 2,f}$, where the SVR predictions exhibit errors approximately $3$ orders of magnitude larger. The superior performance of $k$-NN indicates that the regression quantities in this input space are well approximated by adjacent neighbors, whereas parametric methods such as SVRs struggle to capture the underlying structure. Nevertheless, median relative errors for $M_{\rm 2, f}$ (in cases where the star survives) remain below $2\%$ for the SVR. 

\vspace{0.5cm}
\subsection{Neural Network} \label{sec:reg_NN}

As in Section \ref{sec:class_NN}, the NN used for the regression task is an MLP composed of three hidden layers with $512, 256$, and $128$ neurons, respectively. Each layer is followed by a ReLU activation function and layer normalization. A key architectural feature is the application of a softmax function to the network outputs, which enforces mass conservation by requiring that the three predicted mass fractions are positive and sum to $1$. 

The loss is defined as the MAE of the predictions. Because the predicted values correspond to the final fractional mass in each stellar component and the unbound mass, the loss represents the error in the final predicted masses normalized by the total initial mass of the system. In addition, the contribution to the loss from the predicted unbound mass term is weighted so that it is less than or equal to the loss from the predicted masses. This is done to encourage the model to prioritize minimizing errors in the predicted stellar masses rather than the unbound mass. The value of this weight (\texttt{auxweight}) is treated as a hyperparameter and optimized during tuning. 

\begin{deluxetable}{c | c | c }
    \tabletypesize{\scriptsize}
    \setlength{\tabcolsep}{0.7\tabcolsep}  
    \centering
    \tablecaption{\label{table:reg_params}Regression Hyperparameters}  
    \tablehead{
    	\colhead{Parameter} &
    	\colhead{Optimized Value}  &
        \colhead{Hyperparameter Search Range} } 
    \startdata
    $\rm Epochs$  &  $ 2000 $ & $500 - 2000 $ \\
    $\rm Batch \, size$  &  $ 64 $ & $64 - 512 $ \\  
    $\rm Optimizer$  &  $ \texttt{AdamW} $ & $\texttt{AdamW, sgd}$ \\ 
    $\rm Scheduler$  &  $ \texttt{CA-LR} $ & $\texttt{CA-LR, RLRP}$ \\ 
    $\rm Learning \, rate$  &  $ 0.0002 $ & $10^{-5} - 0.1$ \\  
    $\rm \texttt{patience}$  &  $ \cdots $ & $10 - 120$ \\ 
    $\rm \texttt{factor}$  &  $ \cdots $ & $0.2 - 0.8$ \\  
    $\rm \texttt{auxweight}$ &  $ 0.31 $ & $0.05 - 1.0$ \\  
    \enddata
    \tablecomments{Hyperparameter optimization search for the regression task. Acronyms: stochastic gradient descent (\texttt{sgd}), AdamW optimizer (\texttt{AdamW}), cosine annealing learning rate scheduler (\texttt{CA-LR}), and reduce learning rate on plateau scheduler (\texttt{RLRP}).}
\end{deluxetable}

We perform hyperparameter tuning as described in Section \ref{sec:class_NN} and list the explored ranges for each parameter along with the optimal configuration in Table \ref{table:reg_params}. For each hyperparameter configuration, the epoch with the lowest validation loss is selected to avoid overfitting. The optimized epoch value ($2000$) lies at the edge of the explored range, but we confirmed that training had converged. The model from the sweep with the lowest validation median absolute error in $M_{\rm 1,f}$ is selected as the best-performing model. Median-based metrics are adopted because they more accurately reflect the typical regression error and are less sensitive to outliers arising from misclassified points near decision boundaries. Furthermore, errors in $M_{\rm 1, f}$ were chosen as our primary metric to select the best-performing model for two reasons: ($1$) in cases where only a single star survives, the final mass is always assigned to star $1$, and ($2$) minimizing errors in $M_{\rm 1, f}$ effectively improves performance for the remaining predicted quantities due to the enforced mass conservation restriction. Following the same procedure as in Section~\ref{sec:class_NN}, we conduct an additional sweep that varying the random seed to assess model stability. This yields a best-performing model with median absolute errors of $0.0047 \msun$ for $M_{\rm 1,f} $ and $3.7 \times 10^{-7}$ for $M_{\rm 2,f}$. Errors in $M_{\rm 2,f}$ are smaller than those in $M_{\rm 1,f} $ because $M_{\rm 2,f}$ is zero in a significant portion of the dataset (e.g., in cases of stellar merger or mutual disruption), and the model accurately predicts these zero values. All regression performance metrics are listed in Table~\ref{table:reg_stats}.

The median absolute errors for cases in which star $1$ survives are comparable to the values reported in Table~\ref{table:reg_stats}.
When the final mass of star $2$ is nonzero (label $2$ cases, i.e., flybys) the $k$-NN, SVR, and NN achieve median absolute errors of $0.017, 0.022,$ and $0.0026 \msun$, respectively, with the NN still outperforming all other methods. This trend is also reflected in the relative errors, which include only cases in which the stars survive, as well as in the violin plots shown in Figure~\ref{fig:violin} discussed below.

\begin{figure*}
    \centering
    \includegraphics[width=\linewidth]{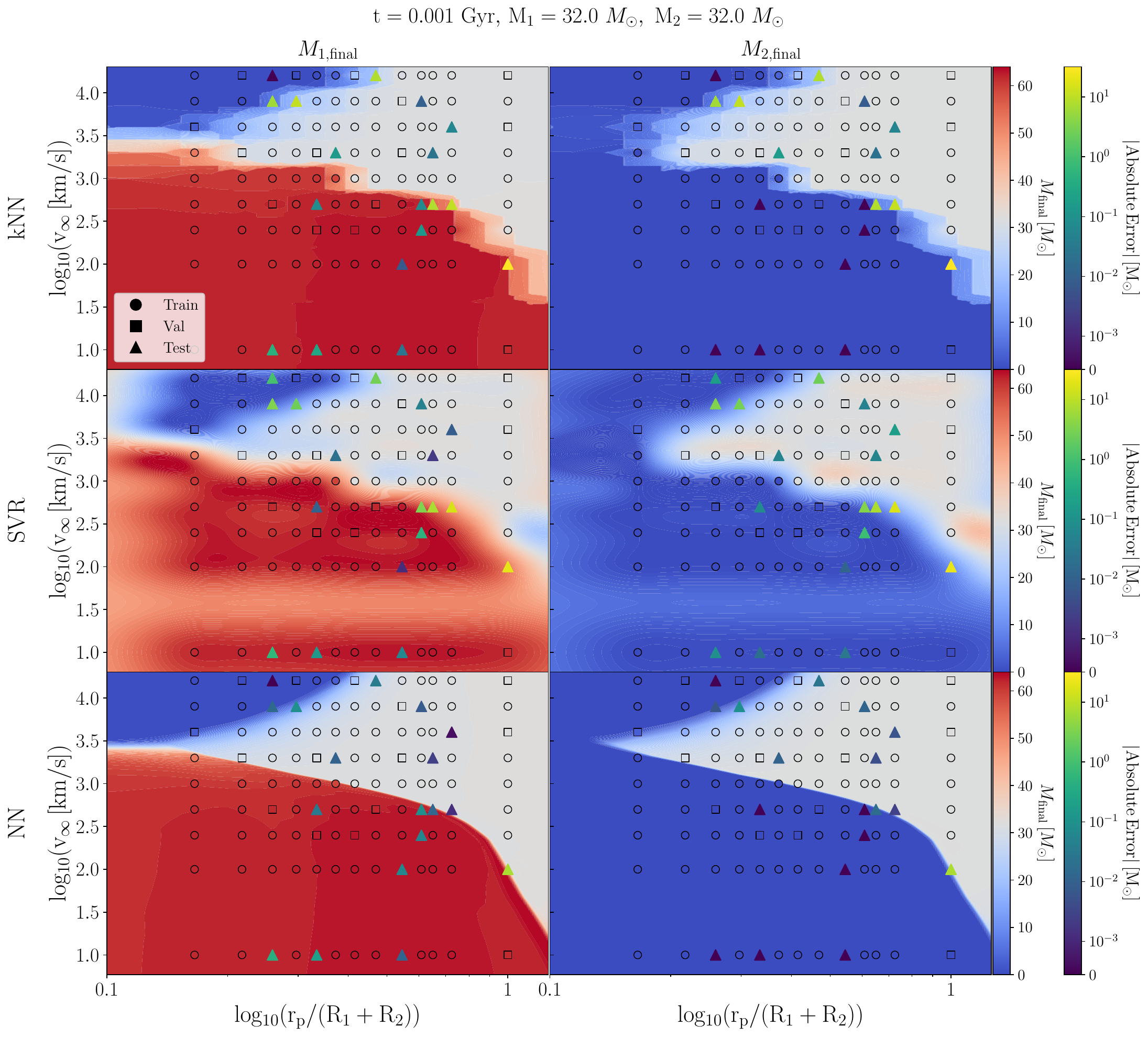}
    \caption{\label{fig:decision_boundaries2} Examples of two-dimensional regression maps for the final stellar masses as functions of pericenter distance ($r_{\rm p}$) and relative velocity at infinity ($v_{\infty}$). From top to bottom, the rows indicate predictions from k-nearest neighbors ($k$-NN), support vector regression (SVR), and neural networks (NN). The first two columns show the predicted final stellar masses using the color gradient, while the rightmost column shows the absolute prediction error for the test dataset. Different markers indicate points from the training (circles), validation (squares), and test (triangles) datasets. The training and validation points are colored according to the ground-truth values.  }
\end{figure*}

\begin{figure*}
    \centering
    \includegraphics[width=\linewidth]{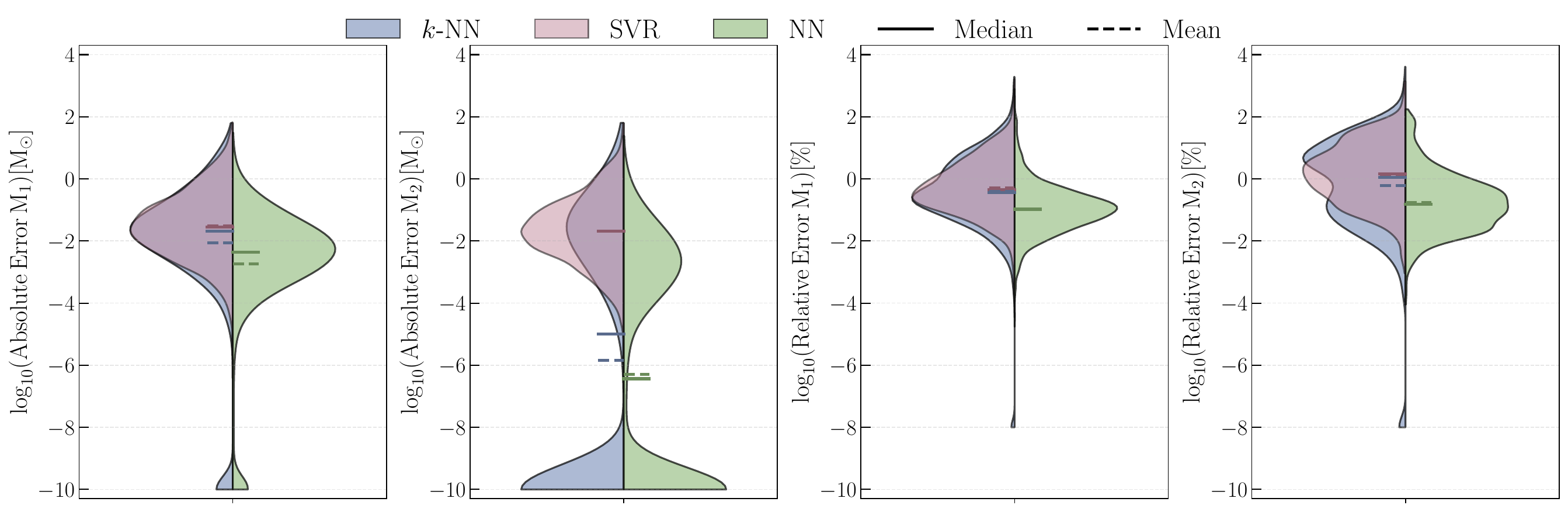}
    \caption{\label{fig:violin} Absolute (left two panels) and relative (right two panels) error distributions for the predicted final stellar masses. Colors indicate results from different algorithms. Colored solid (dashed) bars indicate the median (mean) of each distribution.}
\end{figure*}

Figure \ref{fig:decision_boundaries2} shows the predicted final stellar masses as functions of pericenter distance and relative velocity at infinity for a collision between two $32 \msun$ stars at $1$~Myr. The model predictions are shown as shaded contour maps, while training and validation samples are shown as scatter points colored by their ground-truth values. Test points are colored according to the absolute errors of their predictions. 

Qualitatively, despite its simplicity, the $k$-NN model captures the boundaries between different collision outcomes and exhibits relatively smooth predictions within each region. In contrast, the SVR displays nonphysical features within individual regions, such as localized increases in mass loss at velocities between $10$ and $100$ km/s. The NN accurately recovers the transition boundaries and, like $k$-NN, reproduces characteristic features of equal-mass collisions, such as identical final masses when both survive (shown in light gray).

Quantitatively, the largest errors occur near transition boundaries, as expected. These errors are in fact due to misclassifications and are fundamentally different from regression errors associated with correctly classified data points. For example, errors in data points away from the transition boundaries tend to be smaller. 

To characterize the error distributions across the entire test dataset, Figure~\ref{fig:violin} shows violin plots of the absolute and relative errors for all three methods. For predictions of $M_{\rm 1, f}$, $k$-NN and SVR exhibit similar absolute and relative error distributions, whereas the NN achieves a lower mean and median errors.

For $M_{\rm 2, f}$, the $k$-NN and NN error distributions show pronounced tails toward small absolute errors, corresponding to cases in which the final mass is correctly predicted to be close to zero (i.e., collisions where either only one star survives or both stars are destroyed). Although the SVR exhibits a larger median absolute error for $M_{\rm 2,f}$, its relative error distribution indicates that, in cases where $M_{\rm 2,f}$ is nonzero and therefore relevant, the SVR performs comparably to $k$-NN. The NN achieves median and mean relative errors below $1\%$ for both stellar masses. Overall, the error distributions shown in Figure~\ref{fig:violin}, together with the performance metrics in Table~\ref{table:reg_stats}, demonstrate that the NN achieves the lowest errors across all regression targets and is therefore the best-performing method.

\section{Mixture of Experts}
\label{sec:MoE}

While training a classifier and a regressor separately yields high accuracies, here we investigate whether a shared architecture between the two tasks can further improve model performance. To this end, we draw inspiration from the Mixture of Experts (MoE) architecture \citep{Moe}, which is designed to handle complex data by training different ``experts''  on separate tasks. A gating mechanism, or router, determines which data each expert specializes in, and optimizing this gating mechanism is crucial for achieving strong model performance.

In our approach, we leverage the classification predictions as a gating mechanism to route samples to specialized regression experts, each trained exclusively on data from a single class.
This design offers two key advantages. First, it enhances interpretability, as it is explicitly known which type of data should be routed to each expert, unlike optimized gating mechanisms in which routing decisions are learned implicitly. Second, the architecture is motivated by the strong dependence of the regression properties on collision outcomes. For instance, mutual-destruction events will always result in both final masses being zero. By training individual experts on separate collision outcomes, each model can specialize in distinct physical regimes rather than requiring a single model to capture all regression patterns simultaneously.

\begin{figure*}
    \centering
    \includegraphics[width=0.8\linewidth]{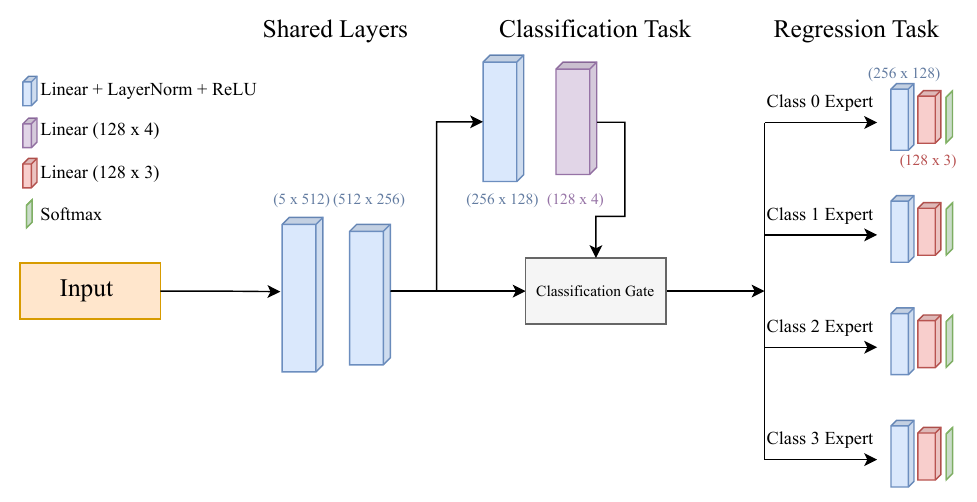}
    \caption{\label{fig:Moe} Diagram of our Mixture of Experts (MoE) neural network. The five-dimensional input data are passed through a shared backbone before being sent to a classification head. The classification labels are used as the gating mechanism to send the data to their specialized experts according to the collision-outcome class. The regression experts share the same NN structure as the classification heads, with an additional soft max activation function applied to ensure mass conservation. }
\end{figure*}

The MoE architecture is shown in Figure~\ref{fig:Moe}. The five-dimensional input data are first passed through a shared backbone consisting of two fully connected layers with $512$ and $256$ neurons, respectively, each followed by layer normalization and ReLU activation. The resulting features are then fed into a classification head composed of a fully connected layer with $128$ neurons with layer normalization and ReLU activation, followed by a final layer that maps to four-dimensional logits corresponding to the collision-outcome classes. The predicted class label is then used as a gating mechanism to route the data to their corresponding regression expert. Each expert uses the same single-layer architecture as the classification head, but we apply a final linear layer that outputs the three predicted values. A softmax activation is then applied to ensure mass conservation. 

When training a multitask NN such as our MoE architecture, the structure of the loss function is crucial for optimizing task-specific metrics. As described in Sections \ref{sec:class_NN} and \ref{sec:reg_NN}, we use cross-entropy loss for the classification task and MAE loss for the regression task. The regression loss comprises of two terms, one for the predicted final mass fractions in each star, and another for the fraction of unbound mass. The weight assigned to the unbound mass term is optimized during hyperparameter tuning (\texttt{auxweight}). To combine the task-specific losses into a joint loss, we use uncertainty-weighted loss \citep[see Equation (10) in][]{Kendall2017}, together with the modification to the regularization term outlined in \cite{Liebel2018}. This approach weights each task-specific loss by a learnable parameter that is jointly optimized with the network weights, eliminating the need to decide the relative weighting of each task in the loss function.

\begin{deluxetable}{c | c | c }
    \tabletypesize{\scriptsize}
    \setlength{\tabcolsep}{0.7\tabcolsep}  
    \centering
    \tablecaption{\label{table:moe_params}MoE Hyperparameters}  
    \tablehead{
    	\colhead{Parameter} &
    	\colhead{Optimized Value}  &
        \colhead{Hyperparameter Search Range} } 
    \startdata
    $\rm Epoch$  &  $ 500 $ & $500 - 2000 $ \\
    $\rm Batch \, size$  &  $ 128 $ & $64 - 512 $ \\ 
    $\rm Learning \, rate$  &  $ 0.0004 $ & $10^{-5} - 0.1$ \\  
    $\rm Optimizer$  &  $ \texttt{AdamW} $ & $\texttt{AdamW, sgd}$ \\ 
    $\rm Scheduler$  &  $ \texttt{CA-LR} $ & $\texttt{CA-LR, RLRP}$ \\ 
    $\rm \texttt{patience}$  &  $ \cdots $ & $5 - 80$ \\ 
    $\rm \texttt{factor}$  &  $ \cdots $ & $0.2 - 0.8$ \\  
    $\rm \texttt{auxweight}$ &  $ 0.77 $ & $0.05 - 1.0$ \\  
    \enddata
    \tablecomments{Hyperparameter optimization search for the regression task. Acronyms: stochastic gradient descent (\texttt{sgd}), AdamW optimizer (\texttt{AdamW}), cosine annealing learning rate scheduler (\texttt{CA-LR}), and reduce learning rate on plateau scheduler (\texttt{RLRP}).}
\end{deluxetable}

A Weights and Biases sweep is performed, with the explored hyperparameter ranges and optimized values listed in Table \ref{table:moe_params}. The epoch with the lowest validation loss is chosen, as a low validation loss corresponds to high balanced accuracy and low absolute errors in the predicted mass fractions. We score models using the metric described in Appendix \ref{sec:model_score}, which prioritizes classification performance while rewarding low regression errors above a defined threshold. As in the previous sections, we perform an additional sweep over different random seeds, yielding a best-performing model with a balanced accuracy in the classification task of \moeBalAccBest $\%$ and median absolute errors of $0.01 \msun$ and $0.000004\msun$ in $M_{\rm 1, f}$ and $M_{\rm 2, f}$, respectively. Additional performance metrics are listed in Table~\ref{table:moe_stats}. 
Note that some of the reported errors are smaller than the mass of a single SPH particle. Nevertheless, they remain informative in this context, as they indicate the errors from the NN predictions.

\begin{deluxetable*}{c | cccccc}
\renewcommand\cellgape{\Gape[3pt]}
\tabletypesize{\scriptsize}
\tablecaption{MoE and Separate NN Performance Metrics\label{table:moe_stats}}
\tablehead{
    \colhead{Model} &
    \colhead{Accuracy} &
    \colhead{Balanced Accuracy} &
    \multicolumn{2}{c}{Median Absolute Error ($\mathrm{M}_\odot$)} &
    \multicolumn{2}{c}{Median Relative Error} \\
    \colhead{} &
    \colhead{(\%)} &
    \colhead{(\%)} &
    \colhead{$M_{1,f}$  [$\times10^{-3}$]} &
    \colhead{$M_{2,f}$  [$\times10^{-3}$]} &
    \colhead{$M_{1,f}$  [$\times10^{-3}$]} &
    \colhead{$M_{2,f}$  [$\times10^{-3}$]}
}
\startdata
MoE &
\makecell{\moeAcc \\ (\text{Best:} \moeAccBest)} 
& \makecell{\moeBalAcc \\ (\text{Best:} \moeBalAccBest)} 
& \makecell{\moeAbsMOne \\ (\text{Best:} \moeAbsMOneBest)} 
& \makecell{\moeAbsMTwo  \\ (\text{Best:} \moeAbsMTwoBest)} 
& \makecell{\moeRelMOne \\ (\text{Best:} \moeRelMOneBest)} 
& \makecell{\moeRelMTwo \\ (\text{Best:} \moeRelMTwoBest)} \\
\hline
Separate NN &
\makecell{\NNAcc \\ (\text{Best:} \NNAccBest)} 
& \makecell{\NNBalAcc \\ (\text{Best:} \NNBalAccBest)} 
& \makecell{\NNAbsMOne \\ (\text{Best:} \NNAbsMOneBest)} 
& \makecell{\NNAbsMTwo  \\ (\text{Best:} \NNAbsMTwoBest)} 
& \makecell{\NNRelMOne \\ (\text{Best:} \NNRelMOneBest)} 
& \makecell{\NNRelMTwo \\ (\text{Best:} \NNRelMTwoBest)} 
\enddata
\tablecomments{Classification and regression performance metrics for the Mixture of Experts (MoE) model. We report the mean and standard deviation across $10$ runs with optimized hyperparameters and different random seeds. Relative errors are computed only for cases in which the star survives. The metrics for the classification and regression neural networks described in Sections~\ref{sec:class_NN} and \ref{sec:reg_NN}, respectively, are included for convenience.  Note that the regression performance metrics are scaled by a factor of $10^{-3}$.}
\end{deluxetable*}

\begin{figure}
    \centering
    \includegraphics[width=0.9\linewidth]{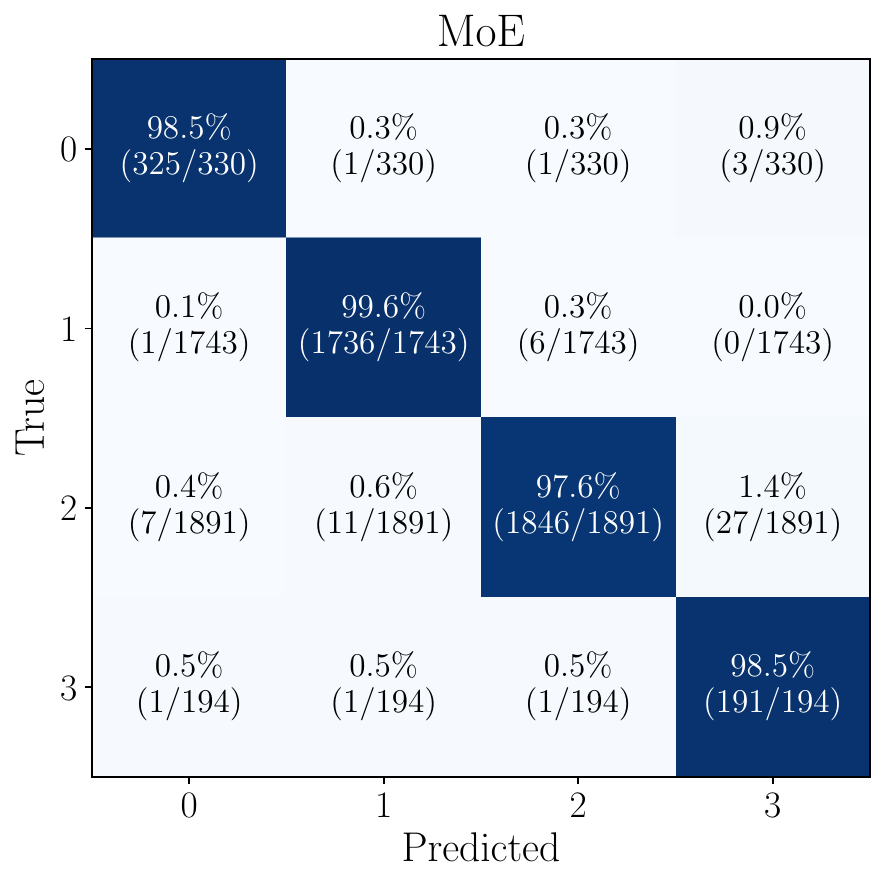}
    \caption{\label{fig:moe_confusion_matrix} Confusion matrix for the test data using the Mixture of Experts (MoE) model. The accuracy and total number of samples in each class are shown. }
    
\end{figure}

\vspace{-0.5cm}
Figure~\ref{fig:moe_confusion_matrix} shows the confusion matrix of the best-performing MoE model. The classification balanced accuracy, also listed in Table~\ref{table:moe_stats}, is comparable to that of the SVR and NN models discussed in Sections~\ref{sec:class_SVM} and \ref{sec:class_NN}, indicating that the gating mechanism is well trained. This is particularly important because the accuracy of the regression predictions relies on the data being routed to the appropriate regression head. 

\begin{figure*}
    \centering
    \includegraphics[width=\linewidth]{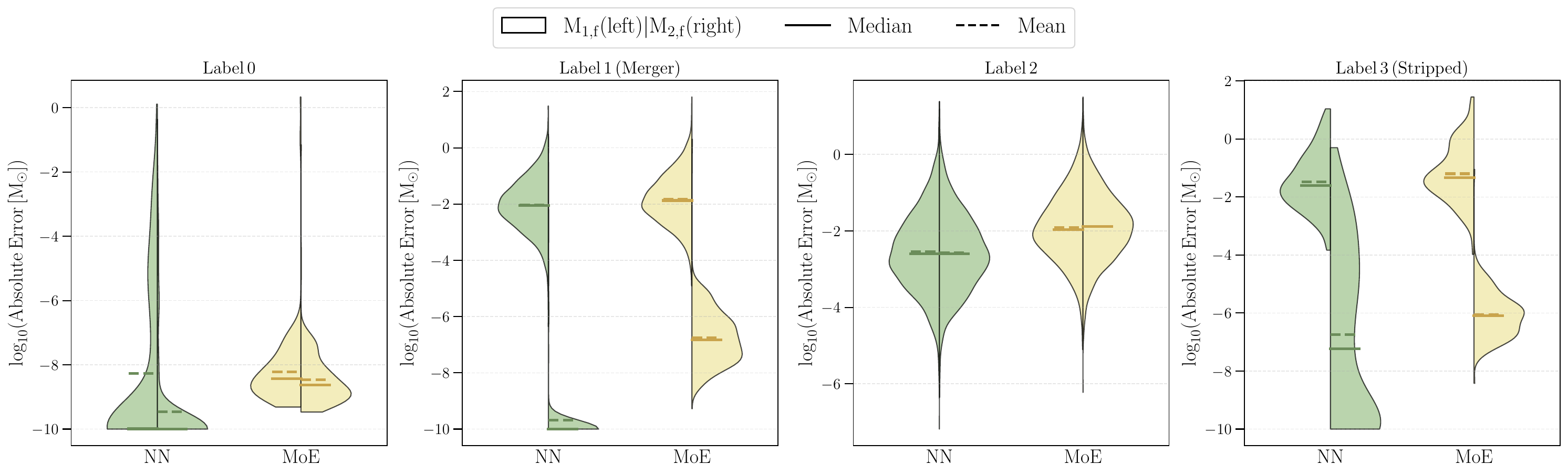}
    \caption{\label{fig:moe_violin} Distributions of absolute errors in the final predicted masses using the neural network (NN) described in Section~\ref{sec:reg_NN} and the Mixture of Experts (MoE) model described in Section~\ref{sec:MoE}. Each panel shows the distribution of absolute errors for the final masses of star $1$ on the left and star $2$ on the right of the violin plot, for data with a specific ground-truth classification label. Small errors are rounded to $10^{-10}$. Solid (dashed) lines indicate the median (mean) values of the raw distribution.}
\end{figure*}

Table~\ref{table:moe_stats} summarizes the performance metrics for the regressed quantities, with the metrics from the individually trained NNs described in Sections~\ref{sec:class_NN} and Sections~\ref{sec:reg_NN} included for convenience. To visualize differences in the error distributions between the individually trained NN and the MoE architecture, Figure~\ref{fig:moe_violin} shows the distributions of absolute errors in the final mass predictions for different ground-truth class labels. Across all labels and for both masses, the regression NN consistently achieves lower mean and median errors compared to the MoE.

For class~$0$ outcomes, both methods yield median and mean errors well below the minimum precision at which the data are physically meaningful. For label~$1$ data, the errors in the final mass of the merger product $M_{\rm 1,f}$ (always assigned to star $1$) are comparable between the two methods, as are the errors for$M_{\rm 1,f}$ in the label~$3$ case. For both final masses in label~$2$ cases, the separately trained NN outperforms the MoE predictions. The effectiveness of the routing gates in reducing high-absolute-error outliers can be seen in the error distributions for $M_{\rm 1,f}$ in case $0$ and $M_{\rm 2, f}$ in case $3$, where the NN exhibits a tail toward higher errors, while the MoE exhibits a more constrained distribution at lower errors.

The observed differences in the error distributions arise from the structure of the loss function and its effectiveness in balancing the relative importance of each task. In the individually trained regression NN, the loss function includes only terms directly related to the regression quantities. In contrast, the MoE loss function combines terms from both the classification and regression tasks adaptively throughout training. Details of the MoE loss function during training are provided in Appendix~\ref{sec:moe_loss}. Future studies will explore alternative loss functions and MoE architectures, including setting the final masses of the corresponding stars equal to zero during training within experts where stars are destroyed, to improve predictive accuracy.

\section{A New software Tool for Stellar Collisions} \label{sec:collAIder}

Building on the models developed in the previous sections, we introduce \texttt{collAIder}~\footnote{https://github.com/elenagonzalez870/collAIder}, a tool designed to predict stellar interaction outcomes and remnant properties. This tool leverages ML models to provide rapid predictions for interactions between two MS stars across a wide range of masses, pericenter distances, and velocities. While prediction accuracy is highest within the parameter space sampled by the SPH grid, the tool has been designed to generalize beyond the direct physical collision regime, extending into the tidal capture and flyby limits. Extrapolation beyond the sampled space can yield inaccurate results, which we explore in Section~\ref{sec:stresstests}.

The largest pericenter distance sampled in our grid is the sum of the individual stellar radii. However, stellar interactions at larger radii can yield interesting results as well. For example, close passages between stars can induce stellar oscillations that dissipate orbital energy \citep[e.g.,][]{PT, Lee1986}. If the tidal perturbation is sufficiently strong, on the order of the orbital energy, the two stars can become gravitationally bound. This process is known as ``tidal capture,'' first invoked to explain the close X-ray binaries observed in globular clusters \citep{Fabian1975}. At distances larger than the capture radius, the stars instead simply pass by each other without forming a bound system or colliding, i.e., a simple flyby. 

The pipeline begins by identifying whether a given set of initial interaction properties are in the collision, tidal capture, or flyby regime. To do this, for a given stellar interaction with $r_p > R_1 + R_2$, we calculate the energy dissipated by tides in the following way. 

\cite{PT} and \cite{Lee1986} derived the expression for tidal energy dissipation as

\begin{equation} \label{eq:Etides}
\Delta E_1 = \left( \frac{GM_{1}^2}{R_{1}} \right) \left( \frac{M_{2}}{M_{1}} \right)^2 \sum_{l = 2}^{\infty} \left( \frac{R_{1}}{r_p} \right)^{2l+2} T_l(\eta)
\end{equation}
where $l=2$ corresponds to the quadrupole term, $l=3$ to the octupole term, etc., and $T_l(\eta)$ is a dimensionless function of 
\begin{equation} \label{eq:eta}
\eta = \left(\frac{M_1}{M_1+M_2}\right)^{\frac{1}{2}} \left(\frac{r_p}{R_1}\right)^{\frac{3}{2}}
\end{equation}

Most of the orbital energy is transferred through the quadrupole and octupole perturbations, so we limit our approximations to only those terms. For simplicity, and to enable the use of predetermined fits, we assign each star an effective polytropic index for the purpose of computing $T_l(\eta)$ values. For stars with $M < 0.4 \msun$ we use $n = 1.5$ (appropriate for fully convective low-mass MS stars), while for stars with $M > 0.8 \msun$ we use $n = 3$ (representing a more centrally condensed structure). This mapping is adopted for computational convenience. A more self-consistent approach would compute $T_l(\eta)$ for each stellar model from linear oscillation calculations, for example using \texttt{GYRE} \citep{Townsend+13,Meng+23}, but this is beyond the scope of the present work. For stars with masses between $0.4$ and $0.8 \msun$, we calculate the energy from tides for both polytropic indices at a given mass, and interpolate the approximate values as

\begin{equation} \label{eq:E_approx}
E_{\rm approx} = \frac{E_3 (M_{\star}-0.4\msun) + E_{1.5} (0.8\msun - M_{\star})}{0.4 \msun}
\end{equation}
where $E_3$ and $E_{1.5}$ are the energy losses due to tidal dissipation for $n=3$ and $n=1.5$ at a given mass $M_{\star}$, respectively. The values of $T_l(\eta)$ are taken from the fits in \cite{Zwart1993}. We do not apply these fits for cases with $\eta>10$, instead assuming that energy dissipation through tidal interactions is negligible (since large $\eta$ corresponds to pericenter distances in the flyby regime). 

Finally, to estimate stellar radii for a given mass, we interpolate from the two nearest stellar mass tracks provided by \texttt{Posydon} v$2$ \citep{Andrews2025}. More specifically, our tool includes an HDF5 file containing single-star \texttt{MESA} hydrogen MS evolution tracks, which finely sample initial stellar masses from $0.1$ to $300 \, \msun$. For a given stellar mass and age, we interpolate each radius to the target age. We then perform a second interpolation between the two radii based on the target mass. This method enables rapid and accurate estimation of stellar radii, using \texttt{MESA} models generated with stellar evolution prescriptions closely matched to those used in the SPH grid. 

These single-star \texttt{MESA} models also enable us to determine whether a given star has evolved beyond the TAMS. For each star, we compare the central hydrogen abundances of the neighboring tracks in mass and age used for interpolation. We define the TAMS as the epoch at which the central hydrogen abundance falls below $10^{-5}$. We then take the minimum of the two TAMS ages and raise a \texttt{ValueError} if the user-provided age exceeds the TAMS by more than $10\%$. This step prevents extrapolation into evolutionary phases where collision outcomes become highly sensitive to stellar structure and are not part of the current training data. 

\begin{figure*}[ht!]
    \centering
    \includegraphics[width=0.8\linewidth]{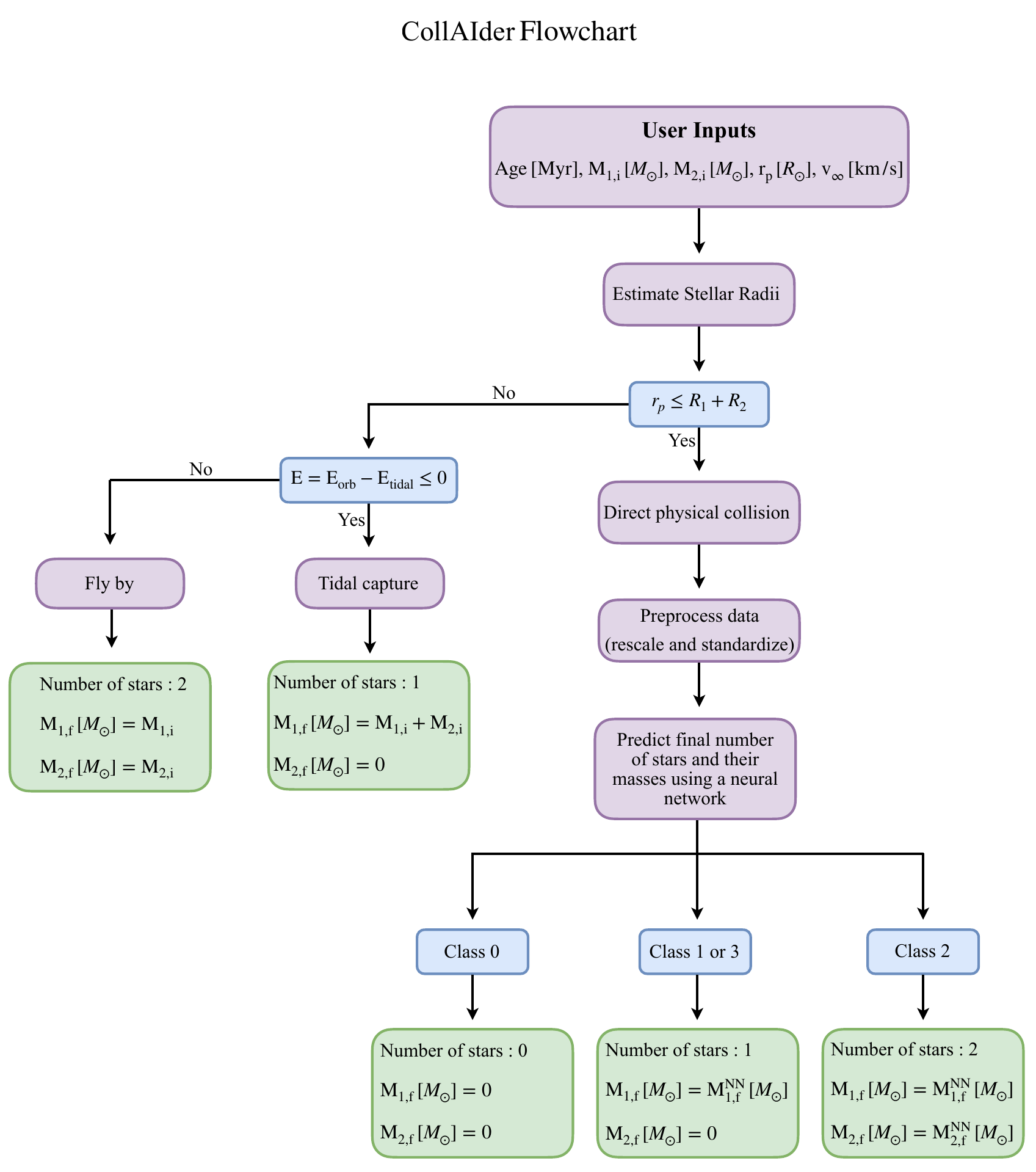}
    \caption{\label{fig:collAIder_flowchart} Flowchart illustrating the decision process in \texttt{collAIder}, described in Section~\ref{sec:collAIder}. Blue boxes indicate decision points, while green boxes denote the final outputs of the software package. The neural network is only used when the input parameters correspond to a physical collision, and the regression outputs are adjusted based on the classification decisions. Masses with a ``NN'' superscript indicate predictions obtained using the neural network. }
\end{figure*}

Finally, if the interaction is determined to be in the direct-collision regime ($r_p \leq R_1 + R_2$), the two ML models are used to predict the outcome and final masses. If the interaction is in the tidal-capture regime, a merger with no mass loss is assumed \citep[e.g.,][]{BenzHills87, Lai1993}. Lastly, in the case of a flyby, both stellar masses are returned unchanged. A flowchart illustrating the decision process in \texttt{collAIder} is shown in Figure~\ref{fig:collAIder_flowchart}.

Given the marginal outperformance of the individually trained NNs over the MoE, we employ the former as the ML model to predict collision outcomes. As shown in the first panel of Figure~\ref{fig:moe_violin}, the regression model returns continuous values and therefore never predicts a final mass exactly equal to zero. Rather than applying a minimum mass cutoff for the remnant object—which makes assumptions about what is considered a collision product—we adjust the regression predictions based on the classification output to ensure consistency. 

For example, when the classifier predicts a merger, the regression model may return a small (but nonzero) mass for star $2$, alongside a correct prediction for the merger remnant. To address this discrepancy, we set $M_{\rm 2,f}$ exactly to zero and add its predicted mass to the ejected mass. This approach is valid because the NN was trained such that the merger remnant mass is always assigned to $M_{\rm 1,f}$. 

Consequently, the performance metrics change compared to those listed in Table~\ref{table:class_stats} and Table~\ref{table:reg_stats}. We report the adjusted performance metrics on the test dataset using \texttt{collAIder} in Table~\ref{table:collAIder_stats} and evaluate its performance in both interpolation and extrapolation in the section below. For the test dataset, we observe a slight decrease in accuracy and balanced accuracy for the classification task. This occurs because some inputs are classified as tidal captures or flybys (instead of direct collisions) due to slight differences between the stellar radii predicted by our \texttt{MESA} models and those from the \texttt{Posydon} v$2$ dataset. Nevertheless, the performance difference is negligible. For the regression task, errors remain comparable. The median error of $0 \msun$ for $M_{\rm 2,f}$ results from the added step of rounding $M_{\rm 2,f}$ to zero in cases of mergers or mutual destruction.

\section{Stress Tests}
\label{sec:stresstests}
The full parameter space for initial conditions is vast and extends beyond the values sampled in this study. To test the robustness of our NN models, we performed additional SPH simulations both within the interpolation regime and in regions outside of the sampled parameter space. 

\begin{deluxetable*}{ccccccc}[t]
\tabletypesize{\scriptsize}
\tablecaption{\texttt{collAIder} Performance Metrics\label{table:collAIder_stats}}
\tablehead{
    \colhead{Dataset} &
    \colhead{Accuracy} &
    \colhead{Balanced Accuracy} &
    \multicolumn{2}{c}{Median Absolute Error ($\mathrm{M}_\odot$)} &
    \multicolumn{2}{c}{Median Relative Error} \\
    \colhead{} &
    \colhead{(\%)} &
    \colhead{(\%)} &
    \colhead{$M_{1,f}$} &
    \colhead{$M_{2,f}$} &
    \colhead{$M_{1,f}$} &
    \colhead{$M_{2,f}$}
}
\startdata
Testing & \collAIderAccTest & \collAIderBalAccTest & \collAIderAbsMOneTest & \collAIderAbsMTwoTest & \collAIderRelMOneTest & \collAIderRelMTwoTest \\
Interpolation & \collAIderAccInt & \collAIderBalAccInt & \collAIderAbsMOneInt & \collAIderAbsMTwoInt & \collAIderRelMOneInt & \collAIderRelMTwoInt \\
TAMS & \collAIderAccTAMS & \collAIderBalAccTAMS & \collAIderAbsMOneTAMS & \collAIderAbsMTwoTAMS & \collAIderRelMOneTAMS & \collAIderRelMTwoTAMS\\
Extrapolation & \collAIderAccExt & \collAIderBalAccExt & \collAIderAbsMOneExt & \collAIderAbsMTwoExt & \collAIderRelMOneExt & \collAIderRelMTwoExt \\
\enddata
\tablecomments{Performance metrics using the \texttt{collAIder} package on the datasets described in Section~\ref{sec:stresstests}. The second and third columns show accuracy and balanced accuracy for the classification task. The fourth and fifth columns show the median absolute errors in the final mass predictions for stars $1$ and $2$, respectively. The sixth and seventh columns show the median relative errors in the final mass predictions for the cases in which the respective star survives.  }
\end{deluxetable*}

\begin{itemize}
    \item \textbf{Interpolation:}
    For the interpolation dataset, we primarily focused on evaluating how well the model generalizes for different initial masses. We perform SPH collisions involving $45$ and $23 \msun$ stars at $2.5 $ Myr, which is roughly halfway through the MS lifetime of the $45 \msun$ star. We ran collisions at pericenter distances enclosing $0, 0.4, 0.8$, and $1$ times the total stellar mass, and at velocities $(10, 500, 2000, 8000)$ km/s. 

    \item \textbf{TAMS Set:} We constructed a grid at $562$ Myr, near the TAMS of a $2 \msun$ star ($\sim 600$ Myr), following the same sampling techniques for secondary masses, velocities, and pericenter distances as the base TAMS grid. Although this grid is almost identical to the existing $2\msun$ TAMS grid, it eas used to test how well the model adapts to collisions of stars near the TAMS. 

    \item \textbf{Extrapolation Set:}
    Crucially, we wanted to assess the reliability of the models on out-of-distribution data. Because our SPH models rely on \texttt{MESA} profiles, we limited this test data set to two stellar masses of $75$ and $90 \msun$, both outside of the range of stellar masses explored in this study. We performed collisions between $75$ and $90 \msun$ stars at $0.07$ and $3.3$ Myr, roughly at the ZAMS of the $75 \msun$ star and the TAMS of the $90 \msun$ star. This set served the purpose of testing whether the algorithm has learned the importance of stellar structure variations over time. We used the same pericenter distances and velocity values as in the interpolation set. 
    
\end{itemize}

The predictions were computed using \texttt{collAIder} to obtain performance metrics representative of those a user would experience. These metrics are summarized in Table~\ref{table:collAIder_stats}. For the classification task, the extrapolation set shows the largest decrease in performance, with a balanced accuracy of \collAIderBalAccExt $\%$. This is expected, since both the interpolation and TAMS datasets lie within the range of explored masses and ages. For the regression task, median absolute errors remain low for most datasets, with the notable exception of the extrapolation dataset, which has a median absolute error of $1.25 \msun$ for $ M_{\rm 1, f}$. However, the corresponding relative error (excluding mutual-destruction cases) is only \collAIderRelMOneExtPer$\%$, indicating good accuracy given that the initial masses are both at least $75 \msun$.  Across all datasets, relative errors in the final mass predictions remain below \collAIderRelMOneExtPer$\%$ and \collAIderRelMTwoExtPer$\%$ for $ M_{\rm 1, f}$ and $ M_{\rm 2, f}$, respectively.

In addition, we performed a targeted stress test motivated by a case from the forthcoming work of Sand et al.\ (2026, in preparation), in which \texttt{collAIder} predicts a small net mass gain by the more massive star during a grazing encounter.  Such outcomes are uncommon but do occur in the SPH dataset. In grazing envelope--envelope interactions, some shocked material is decelerated in the COM frame and may later be reaccreted; if more material originating from star~2 falls back onto star~1 than is lost from star~1, then star~1 can experience a small net mass gain.  Because our regression model predicts the partition of bound mass among the two survivors while enforcing overall mass conservation, it naturally permits this behavior rather than imposing mass loss for each star.

Our specific test involves the collision at age $0.79$~Gyr with $(M_{1,i},M_{2,i})=(1.593,1.128)\,\msun$, $r_p=0.568\,R_\odot$, and $v_\infty=584$~km~s$^{-1}$.  The SPH simulation yields a two-star outcome with $(M_{1,f},M_{2,f})=(1.599,1.104)\,\msun$.  In this case, star~1 gains $0.006\,\msun$ ($0.4\%$) while the system as a whole loses $0.017\,\msun$ ($0.6\%$) to unbound ejecta.  The \texttt{collAIder} prediction for the same initial conditions correctly captures the sign of the effect ($M_{1,f}>M_{1,i}$). Although it overpredicts the accreted mass, the predicted $M_{1,f}= 1.63\,\msun$ remains within $2\%$ of the SPH value.  This demonstrates that the model can reproduce these rare events at a level consistent with the expected accuracy of the regression task.

\section{Discussion and Conclusions}
\label{sec:discussion}
In this work, we have presented a new set of $27{,}720$ SPH simulations of stellar collisions spanning a wide range of ages, masses, relative velocities, and pericenter distances. 
This comprehensive dataset is used to train ML models to predict not only the outcome of the interaction but also the final masses of the remnant stars. In this framework, there are four possible classification labels: $0, 1, 2,$ and $3$. The first three intuitively indicate the number of remnant stars: $0$ denotes mutual-destruction events, $1$ denotes mergers, and $2$ denotes mild collisions. Label $3$ indicates stripped-star cases, where a highly energetic event (at high relative speeds and moderately small pericenter distances) completely destroys one star and partially strips the other. Although one star survives in this case, it is treated as a distinct class since the hydrodynamic evolution of the collision differs significantly from the other outcomes. For the regression task, the model is trained to predict the final fractional mass in each star and unbound material. The NN is structured to ensure mass conservation. We compare the classification and regression performance of three ML algorithms ($k$-NN, SVM, and NNs), finding the following:

\begin{itemize}
    \item \textbf{Classification:} the SVM and NN achieve comparable performance, with balanced accuracies of \svcBalAcc$\%$ and \NNBalAccBest$\%$, respectively. This is unsurprising, as SVMs are well suited for multiclass classification problems. Furthermore, most misclassified data points lie near the decision boundaries, as expected.  

    \item \textbf{Regression:} the NN outperforms all other methods, with median absolute errors in stellar masses of $0.00441\msun$ for $M_{\rm 1,f}$ and $3.7 \times 10^{-7}\msun$ for $M_{\rm 2,f}$. The corresponding relative errors are \NNRelMOneBestPer$\%$ and \NNRelMTwoBestPer$\%$. As is the case for the classification task, most errors occur near decision boundaries, where small deviations in initial conditions produce large changes in the final stellar masses. 
\end{itemize}

We also investigated whether a MoE architecture, consisting of initial shared layers followed by four separate regression experts trained on data corresponding to predicted classification labels, can outperform the separately trained NNs. We find that the MoE achieves comparable balanced accuracy in the classification task and slightly larger errors in regression. Nevertheless, the gating mechanism is well trained and the performance of the MoE model remains comparable to that of the individually trained NNs. Future studies will investigate different MoE architectures to improve performance further. 

Finally, we present the package \texttt{collAIder}, which uses the trained ML models to predict stellar collision outcomes and remnant properties. The package, outlined in Section~\ref{sec:collAIder}, classifies stellar encounters into direct physical collision, tidal capture, and flyby regimes, and returns predictions for a wide range of dynamical properties. Its performance is tested on interpolation, extrapolation, and near-TAMS cases. The model struggles the most with the extrapolation dataset, yielding a balanced accuracy of \collAIderBalAccExt$\%$ and median relative errors of \collAIderRelMOneExtPer $\%$ and \collAIderRelMTwoExtPer $\%$ in the final predicted masses of stars $1$ and $2$, respectively.

Although already quite extensive, the SPH grid used here to train the ML models encompasses a limited range of sampled stellar models and collision parameters. As a result, performance will inevitably decrease for data outside the sampled ranges. Nevertheless, the extrapolation tests outlined in Section~\ref{sec:stresstests} already show satisfactory performance for collisions involving MS stars with metallicity of $0.01Z_{\odot}$.

The \texttt{collAIder} pipeline approximates stellar radii to determine whether an encounter lies in the direct physical collision regime (in which case the ML model is used for inference), the tidal-capture regime (in which case a merger with no mass loss is assumed), or the distant flyby regime. These stellar radii are interpolated using the \texttt{Posydon} v$2$ \texttt{MESA} tracks \citep{Andrews2025}. While these tracks sample stellar masses and ages very finely, interpolation is still needed to infer radii at arbitrary user inputs.  Uncertainties in these interpolated values introduce small errors in the stellar radii, which can in edge cases cause misclassifications in the nature of the encounter. 

Furthermore, predictions are currently limited to MS stars with metallicity $Z = 0.01 Z_\odot$. Future work will expand into other metallicities as well as incorporate collisions involving giant stars and compact objects. 

This work showcases the potential impact and applicability of ML algorithms trained on large datasets. Such NN models can dramatically reduce computational costs while delivering physically consistent results almost instantaneously. A particularly impactful application lies in $N$-body  simulations, where no mass loss is often assumed for stellar collisions. Incorporating ML models into large-scale $N$-body frameworks will enable more realistic treatments of stellar collisions, with crucial implications for understanding collision remnants and their properties, including blue stragglers, black holes in the upper-mass gap, and the transient signals that may accompany these collisions. Forthcoming work (Sand et al.\ 2026, in preparation) performs initial tests of the NN models presented in this paper by implementing them into semi-analytical models of the Milky Way's Galactic center.

\section{Acknowledgements} 
We thank Ugur Demir, Nabeel Rehemtulla, Ved Shah, and Philipp Srivastava for useful discussions.
This work was supported by NSF grant AST-2511543 to F.A.R.\ and T.S.\ at Northwestern University.
Support for E.G.P.\ was provided by the NSF Graduate Research Fellowship Program under grant DGE-2234667. S.C.R.\ is grateful for support from the Lindheimer Fellowship. F.K.\ and C.E.O.\ acknowledge support from a CIERA Postdoctoral Fellowship. T.C.P.\ was supported in part by NSF grants AST-2149425 and AST-2446392. 
We gratefully acknowledge the support of the NSF-Simons AI-Institute for the Sky (SkAI) via grants NSF AST-2421845 and Simons Foundation MPS-AI-00010513. This work used Bridges-2 at the Pittsburgh Supercomputing Center through allocation PHY-240311 from the Advanced Cyberinfrastructure Coordination Ecosystem: Services \& Support (ACCESS) program, which is supported by NSF grants 2138259, 2138286, 2138307, 2137603, and 2138296.
This research was also supported in part through the computational resources and staff contributions provided for the Quest high-performance computing facility at Northwestern University, which is jointly supported by the Office of the Provost, the Office for Research, and Northwestern University Information Technology. A.F. \ and S.C. \ also acknowledge the support of the Aspen Center for PHYics, funded by the NSF grant PHY-2210452, where some of this work was initiated.

\bibliographystyle{aasjournal}
\bibliography{main}

\appendix

\section{Input Features}
\label{sec:input_features}
\begin{figure*}[h]
    \centering
    \includegraphics[width=\linewidth]{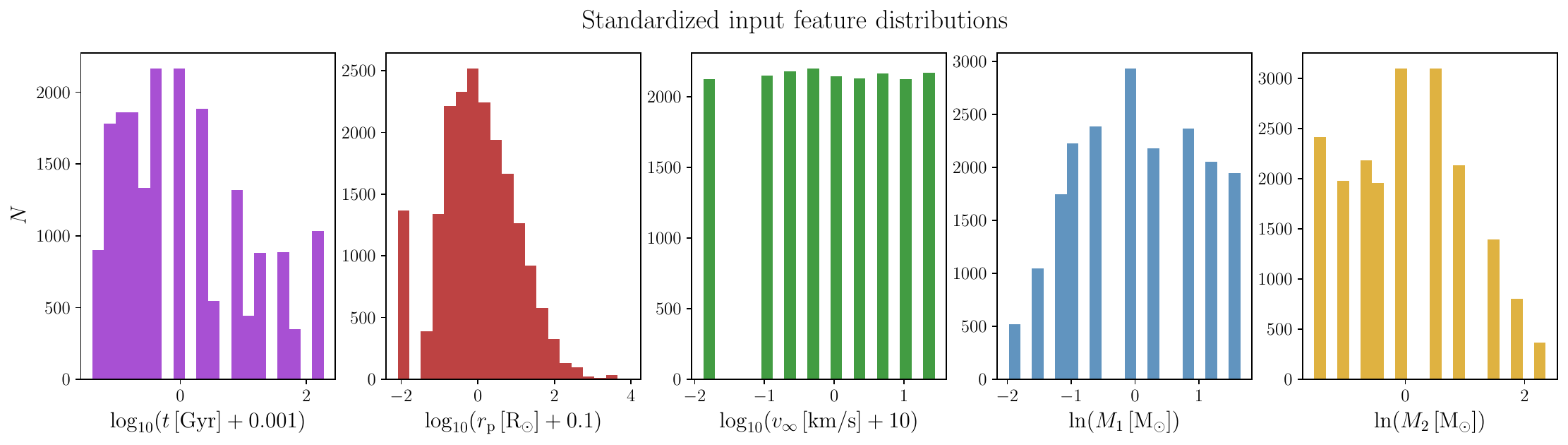}
    \caption{\label{fig:input_features} 
    Distributions of the standardized input features described in Section~\ref{sec:classification}. Note that the axes show the standardized values, and are thus not representative of the physical ranges sampled in the dataset.}
\end{figure*}

The machine learning methods are trained on a five-dimensional input space, consisting of the 
ages ($t$) and masses of the colliding stars ($M_1$ and $M_2$), as well as the kinematic properties of the encounter: the pericenter distance ($r_\mathrm{p}$) and relative velocity at infinity ($v_{\infty}$). Prior to training, the input features are log-transformed (with variable-dependent offsets to avoid singularities near zero) and standardized using Equation~\ref{eq:norm}. The resulting standardized input feature distributions are shown in Figure~\ref{fig:input_features}.

\section{SVR Grid Search}\label{sec:svr_gridsearch}

In this section, we detail the grid search performed for the regression task using the support vector regression (SVR) module in \texttt{scikit-learn}. The grid search is performed for each regressed quantity separately, as summarized in Table~\ref{table:svr_gridsearch}. Three separate kernels were explored: radial basis function (\texttt{rbf} ), polynomial, and sigmoid. A more limited exploration of the polynomial kernel was carried out, since training a single model can take several hours. The hyperparameter search ranges were chosen according to computational resources and to maximize exploration of the parameter space. For all quantities except the degree, we searched over logarithmically spaced values. The parameter \texttt{C} parameter acts as a regularization parameter that controls the complexity of the model, with larger values imposing stronger penalties and leading to more complex models. 
The parameter \texttt{gamma} determines the influence of individual training points and $\epsilon$ defines the width of the tube within which no penalty is assigned to errors. Following an initial coarse grid search, we conducted a refined search around smaller epsilon values using the optimized values found for the kernel, \texttt{C}, and \texttt{gamma}. The final optimized values are listed in Table~\ref{table:svr_gridsearch}. 

\begin{deluxetable}{l|ccccc}[h]
\tabletypesize{\scriptsize}
\setlength{\tabcolsep}{0.7\tabcolsep}  
\centering
\tablecaption{SVR Hyperparameter Grid Search Ranges\label{table:svr_gridsearch}}  
\tablehead{
\colhead{Quantity} &
    \colhead{Kernel} &
    \colhead{\texttt{C}} &
    \colhead{\texttt{gamma}} &
    \colhead{$\epsilon$} &
    \colhead{Degree}} 
\startdata
\hline
\multicolumn{6}{c}{\textbf{Initial Coarse Grid Search}} \\
\hline
$q_{1,f}$ & \texttt{rbf}  & $0.1-100$ & $0.01 - 10$ & $0.001 - 0.1$ & --- \\
$q_{1,f}$ & Polynomial & $0.1-10$ & $0.01 - 1$ & $0.01 - 0.1$ & $2-3$ \\
$q_{1,f}$ & Sigmoid & $0.1-1000$ & $0.01 - 10$ & $0.0001 - 0.1$ & --- \\
\hline
$q_{2,f}$ & \texttt{rbf}  & $0.1-100$ & $0.01 - 10$ & $0.001 - 0.1$ & --- \\
$q_{2,f}$ & Polynomial & $0.1-10$ & $0.01 - 1$ & $0.001 - 0.1$ & $2-3$ \\
$q_{2,f}$ & Sigmoid & $0.1-1000$ & $0.01 - 10$ & $0.0001 - 0.1$ & --- \\
\hline
$q_{\rm{u},f}$ & \texttt{rbf}  & $0.1-100$ & $0.01 - 10$ & $0.001 - 0.1$ & --- \\
$q_{\rm{u},f}$ & Polynomial & $0.1-10$ & $0.01 - 1$ & $0.001- 0.1$ & $2-3$ \\
$q_{\rm{u},f}$ & Sigmoid & $0.1-1000$ & $0.01 - 10$ & $0.0001 - 0.1$ & --- \\
\hline
\multicolumn{6}{c}{\textbf{Refined Fine Grid Search}} \\
\hline
All & \texttt{rbf}  & $1$ & $10, 50$ & $0.00001 - 0.001$ & --- \\
\hline
\multicolumn{6}{c}{\textbf{Optimized Values}} \\
\hline
All & \texttt{rbf}  & $1$ & $10$ & $0.00001$ & --- \\
\hline
\enddata
\tablecomments{Support vector regression (SVR) hyperparameter search for the final fractional masses of star $1$ ($q_{1,f}$) and star $2$ ($q_{2,f}$), and unbound mass ($q_{\rm{u},f}$). A search was conducted across three kernels: the radial basis function (\texttt{rbf} ), polynomial, and sigmoid.}
\end{deluxetable}

\section{MoE Loss Function}\label{sec:moe_loss}

\begin{figure}[h!]
    \centering
    \includegraphics[width=\linewidth]{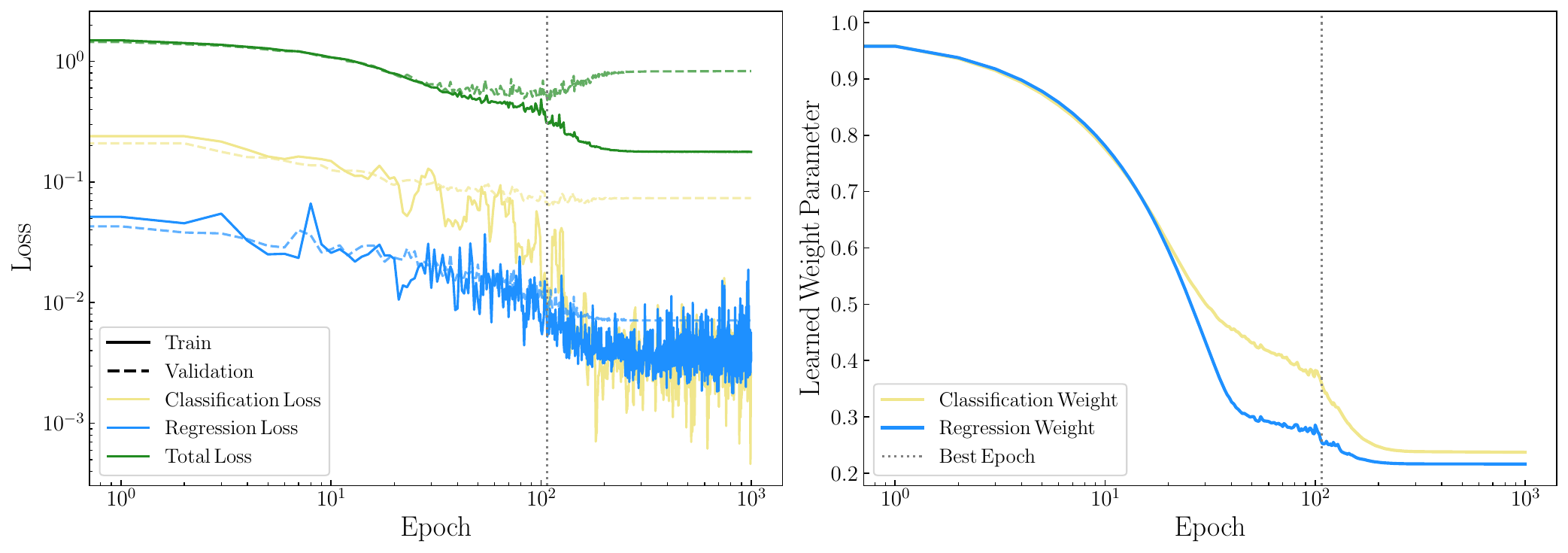}
    \caption{\label{fig:moe_loss} Training (validation) loss functions for the classification and regression tasks shown in solid (dashed) yellow and blue, respectively. The total loss function is shown in green, obtained using uncertainty-weighted loss \citep[][Equation(10)]{Kendall2017}. The classification loss function is smoothed with a moving average of size 5 for visualization purposes. The dashed line indicates the epoch with the lowest validation loss. }
    
\end{figure}

The challenge of any multitask architecture is optimizing performance across all tasks. In this work, we have used uncertainty-weighted loss \citep[see Eq.\ 10 in][]{Kendall2017} with the modification to the regularization term outlined in \cite{Liebel2018}. This method weights the loss of each task by learnable parameters that are optimized jointly with the network weights. In the left panel of Figure~\ref{fig:moe_loss} we show the respective loss functions for the classification and regression tasks, alongside the overall loss of the model during the training. The figure shows both training and validation losses, illustrating that the model begins to overfit after approximately epoch $107$ (kept as the best model), shown by the increase in the validation loss. The monotonic decrease of both task-specific loss functions demonstrates that the uncertainty-weighted approach successfully balances optimization across both tasks. The evolution of the learned weighting parameters is shown in the right panel, with the optimal classification and regression uncertainty weights found to be $0.36$ and $0.26$, respectively.

\vspace{1cm}

\section{Model Scoring}\label{sec:model_score}

Because multitask NNs optimize multiple tasks simultaneously, a metric tailored to the specific science case is needed to evaluate model performance. We score the MoE models using 
\begin{equation} \label{eq:score}
\text{Score} = \alpha' (1 - \text{BA}) + \beta' \, \text{MedAE}_{1}^* + \eta' \, \text{MedAE}_{2}^*
\end{equation}
where BA is the balanced accuracy and $\text{MedAE}_{i}^{*} = (\text{MedAE}_i - \text{MedAE}_{i, \rm min})/(\text{MedAE}_{i, \rm max} - \text{MedAE}_{i, \rm min})$ are the min-max normalized median absolute errors in the predicted masses, with $i = 1,2$ corresponding to stars 1 and 2. The prefactors that weight each component are set to 
\begin{equation}
(\alpha', \beta', \eta') = \begin{cases}
(\alpha + \beta + \eta, 0, 0) & \text{if } \text{MedAE}_1, \text{MedAE}_2 < 0.005\\
(\alpha , \beta + \eta, 0) & \text{if } \text{MedAE}_2 < 0.005\\
(\alpha , 0, \eta + \beta) & \text{if } \text{MedAE}_1 < 0.005 \\
(\alpha, \beta, \eta) & \text{otherwise}
\end{cases}
\end{equation}
where $\alpha = 0.8$, and $\beta = \eta = 0.1$. The value of $\alpha$ is large to prioritize classification performance. Furthermore, we impose a threshold of $0.005 \msun$ on the MedAE, below which we do not reward regression performance. This avoids rewarding improvement past an absolute error of $0.005 \msun$, which is not physically significant at the sampled mass ranges. The model with the lowest score is selected as the best-performing model. Note that this metric is particularly suited for the application of this tool, and other scientific goals might require a different metric to identify the best model.

\end{document}